\documentclass[aip,pof,10pt]{revtex4-1}
\usepackage{multirow}
\usepackage[table]{xcolor}
\usepackage{rotating}
\usepackage{amsmath}
\usepackage{bm}
\usepackage{array}
\usepackage{graphicx}
\usepackage{dcolumn}   
\usepackage{amssymb}
\usepackage[english]{babel}

\begin{document}
\title{Slender-ribbon theory}
\date{\today}
\author{Lyndon Koens\footnote{lmk42@cam.ac.uk}, and Eric Lauga\footnote{e.lauga@damtp.cam.ac.uk}}
\affiliation{ Department of Applied Mathematics and Theoretical Physics, University of Cambridge, Wilberforce Road, Cambridge CB3 0WA, United Kingdom}
\begin{abstract} 

Ribbons are long narrow strips possessing three distinct material length scales (thickness, width, and length) which allow them to produce unique shapes unobtainable by wires or filaments.  For example when a ribbon has half a twist and is bent into a circle it produces a M\"obius strip. Significant effort has gone into determining the structural shapes of ribbons but less is know about their behavior in viscous fluids. In this paper we determine, asymptotically, the leading-order hydrodynamic behavior of a slender ribbon in Stokes flows. The derivation, reminiscent of  slender-body theory for filaments,  assumes that the length of the ribbon is much larger than its width, which itself is much larger than its thickness. The final result is an integral equation for the force density on a mathematical ruled surface, termed the  ribbon plane, located inside the ribbon. A numerical implementation of our derivation shows good agreement with the known hydrodynamics of long flat ellipsoids, and successfully captures the swimming behavior of artificial microscopic swimmers recently explored experimentally. We also study the asymptotic behavior of a ribbon bent into a helix, that of a twisted ellipsoid, and we investigate how accurately the hydrodynamics of a ribbon can be effectively captured by that of a slender filament. Our asymptotic results provide the  fundamental framework necessary to predict the behavior of slender ribbons at low Reynolds numbers in a variety of biological and engineering problems. 
\end{abstract}
\maketitle
\def\v{\vspace{2cm}}

\section{Introduction}

Ribbons are everywhere. Some plants create rigid ribbon-like seed pods to encourage seed  dispersal by the wind \cite{YoForterre} while many  marine animals swim by sending waves down ribbon-like appendages \cite{Lighthill2006,Shirgaonkar2008}. The three dimensional folding structure of proteins can be simplified and understood well using ribbons \cite{Richardson2002} while the super-coiling behavior of DNA has been related to a linking number of ribbons \cite{Fuller1971}.  For years machines have used closed ribbons as drive belts, in order to transfer the power from a motor to other turning objects. Recently, ribbons have also been used to create magnetically-driven artificial swimmers at the micron scale \cite{Zhang2009,Zhang2010,Diller2014}.

The  ability for a ribbon  to take so many configurations comes form the fact that it is characterized by three material length scales: the centerline length, $2\ell$, the plane width, $2b$, and the thickness, $2a$. These  three length scales make the behavior of a ribbon fully three dimensional and allows for the creation of different shapes with complex topology. 
For example a M\"obius strip \cite{Starostin2007}, a looped ribbon with a half twist in it, is very different from a ribbon bent into a loop without any twisting. The ability to make such configurations explains why ribbons are relevant to so many fields of research \cite{YoForterre,Lighthill2006,Shirgaonkar2008,Richardson2002,Fuller1971,Zhang2009,Zhang2010,Eloy2011,Diller2014,Kohno2015}. 

Extensive past work has gone into determining theoretically  the equilibrium structures of ribbons as solid mechanical objects \cite{Starostin2007,Dias2014}, experiments have looked into how ribbons curl \cite{Arriagada2014}, theory has addressed how ribbons let animals swim \cite{Lighthill2006,Shirgaonkar2008} and flags flap \cite{Eloy2011}. However, very little theoretical investigation has gone into the general hydrodynamics of ribbons, particularly at low Reynolds number \cite{Makino2005a,Keaveny2013,Keaveny2011}. The existing computational framework can tackle  ribbons with a width on the same order as its thickness  \cite{Keaveny2013,Keaveny2011}, which are thus  too thick to represent the ribbons seen naturally or used in micro-swimming experiments \cite{Zhang2009,Zhang2010}.

In this paper we derive the asymptotic framework necessary to quantify the hydrodynamics of a thin slender ribbon at low Reynolds number. The slenderness of the ribbon is assumed to be characterized by the asymptotic limit $\ell\gg b \gg a$. An expansion similar to the slender-body theory (SBT) expansion for Stokes flow \cite{Johnson1979} is performed. Using only force singularities (stokeslets), we are able to fully determine the hydrodynamics of the ribbon at leading order in $b/\ell$. This expansion assumes that the curvature of the ribbon and the rate at which the ribbon twists is less than $\ell/b$, similar to the curvature restriction in SBT, and that the ribbon is a ruled surface. The resulting formulation is tested against known solutions and experimental results, and we obtain excellent agreement. We also investigate the behavior of a slender ribbon twisted around its central axis and that of a slender ribbon with a helical centerline. The behavior seen is finally compared with the results predicted by SBT for filaments in order to illustrate the differences in the dynamics of wires and ribbons \cite{Johnson1979}. While the paper focuses solely on rigid-body motion, the slender-ribbon equations can also be applied to deforming ribbons.

The paper is organized as follows.  In Sec.~\ref{sec:stokesflow} we discuss the low-Reynolds number hydrodynamic framework for this study, and give a quick overview of the history and derivation of SBT. Section~\ref{sec:SREderivation} is the main technical part of the paper which presents the derivation of the slender-ribbon theory (SRT) equations. This section starts by describing the mathematical structure of the ribbons considered (Sec.~\ref{sec:ribbondescription}) and then gives an outline to the expansion process (Sec.~\ref{sec:outline}). We then consider in detail how the system should be expanded (Sec.~\ref{sec:expandpoints}), before performing the various expansions (Sec.~\ref{sec: expansion}) and determining the final set of equations (Sec.~\ref{sec:SRE}). We   conclude this section by comparing the resulting equations to the ones obtained for slender filaments (Sec.~\ref{sec:SRTvsSBT}). In Sec.~\ref{sec:numerics} we then describe the numerical procedure used to solve the integral equations arising in the SRT formulation. The working code is then tested against the known analytical formulae  for a long-flat ellipsoid (Sec.~\ref{sec:plateellipsoid}) before comparing the behavior of an asymptotically thin ribbon with a helical centerline to the thicker ribbons recently explored numerically   \cite{Keaveny2013,Keaveny2011} (Sec.~\ref{sec:keaveny}). With the same parametrization, we use the SRT results to address the swimming dynamics of so-called artificial bacterial flagella recently proposed and tested experimentally on microscopic scales (Sec.~\ref{sec:zhang}). We then address how different the results for slender ribbons are from slender bodies, showing in particular that no slender body can effectively capture the hydrodynamics of a helical slender ribbon (Sec.~\ref{sec:sbtequiv}).  Finally the hydrodynamics of  twisted ribbons with straight centerlines is explored in Sec.~\ref{sec:twistellipse}.

\section{Stokes flow and slender-body theory} \label{sec:stokesflow}

This paper intends to determine the hydrodynamic forces on slender ribbons at low Reynolds numbers, with an eye on application to microscopic and biological bodies.  In this case the fluid is accurately described by the incompressible Stokes equations
\begin{eqnarray}
\nabla p  &=&   \mu \nabla^{2} \mathbf{u}, \\
\nabla \cdot \mathbf{u} &=& 0,
\end{eqnarray}
where $\mu$ is the dynamic viscosity of the fluid, $p$ is the pressure and $\mathbf{u}$ is the velocity field. These equations are linear and independent of time.
Therefore, the net hydrodynamic force and torque on a submerged rigid body is linearly related to the linear and angular velocity of the body by
\begin{equation} \label{resistance}
\mathbf{\tilde{F}}  = -\mathbf{R}_{\mathbf{\tilde{F}}\mathbf{\tilde{U}}} \mathbf{\tilde{U}} ,
\end{equation} 
where $\mathbf{\tilde{F}}$  is a six-component vector containing all components of the hydrodynamic forces and torques on the body and $\mathbf{\tilde{U}}$ is a six-component vector containing the  instantaneous linear and angular velocities of the body. 
In Eq.~\eqref{resistance}, $\mathbf{R}_{\mathbf{\tilde{F}}\mathbf{\tilde{U}}}$ is the $6\times6$   resistance matrix which is proportional to the viscosity and only depends on the size and shape of the body.

Computing the resistance matrix is in general not an easy task for an arbitrarily-shaped body and is typically obtained numerically through the use of flow singularities, starting with the Green's function for Stokes equation's.
The flow for this Green's function is given by
\begin{eqnarray}
8 \pi \mu \mathbf{U^{s}}(\mathbf{R};\mathbf{f}) &=& \frac{\mathbf{I} + \mathbf{\hat{R}} \mathbf{\hat{R}}}{|\mathbf{R}|} \cdot \mathbf{f},
\end{eqnarray}
where $\mathbf{R}$ is the vector from the singularity to the desired point of flow, $\mathbf{\hat{R}}$ is the unit vector in the direction of $\mathbf{R}$, and $\mathbf{I}$ is the identity tensor.
This fundamental singularity is called the stokeslet and represents the flow created by a point force in the fluid of strength $\mathbf{f}$. Other singularity solutions to the Stokes equation can be obtained  by taking  derivatives of the stokeslet. For example a potential `source dipole' of strength $\mathbf{g}_{SD}$ is defined as $-1/2$ the Lapacian of the stokeslet and has the form \cite{Chwang2006}
\begin{eqnarray}
8 \pi \mu \mathbf{U^{sd}}(\mathbf{R};\mathbf{g}_{SD}) &=& - \frac{\mathbf{I} - 3 \mathbf{\hat{R}}  \mathbf{\hat{R}}}{|\mathbf{R}|^{3}} \cdot \mathbf{g}_{SD}.
\end{eqnarray}

These singularities can be used to determine the hydrodynamics of any body moving in a Stokes fluid in two ways. In the first method, the Green's function nature of the stokeslet can be used to turn the Stokes equations into integral equations over the surface of the body. This is    the boundary integral method \cite{Pozrikidis1992}. The second method consists of placing stokeslet singularities, and its derivatives, within the body and the use of the boundary conditions on the surface of the body to determine their strengths \cite{Chwang2006}.  This method arises from the linearity of the Stokes equations \cite{Kim2005} and is sometimes called the representation by fundamental singularities. For example the Stokes flow around a rigid sphere of radius $a$ and velocity $U$ is described by a stokeslet and source dipole of strengths $6 \pi \mu a U$ and $-\pi a^3 U$, located at the centre of the sphere \cite{1976,Kim2005}.
 
The distribution of singularities in second method is only known exactly for some simple shapes \cite{Chwang2006} and so for most calculations the dynamics of a body are typically found using boundary integrals. The boundary integral method is very powerful but for some shapes, such as those characterized by a large range of length scales, the surface discretization can be difficult, requiring a fine mesh of the surface and potentially creating long computation times. 
 For long thin bodies with a circular cross section an approximation is typically used,  based on the representation by fundamental singularities, called slender-body theory. 

Slender-body theory (SBT) aims to capture the hydrodynamics of a long thin body by placing stokeslets and source dipoles along its centerline. These singularities are then expanded in two domains: an outer region where the centerline length dominates, and the body has effectively zero thickness, and an inner region where the thickness of the body dominates. The two domains are then  matched asymptotically to determine the flow at the surface of the body assuming that the velocity at the  surface of the body moves rigidly with the centerline. The relative strengths of the source dipoles are then found by ensuring that at each cross section there is no velocity variation across the surface. 

There have been a few different formulations of SBT \cite{Cox,Sol1976,1976,Johnson1979}. Early work used the flow past an infinite cylinder as the inner region and matched the results to a line of stokeslets in the outer region \cite{Cox,Sol1976}. These formulations were only applicable far from the ends of the long thin body and produced a series in powers of $1/\log(\epsilon)$, where $\epsilon$ is the body thickness divided by the centerline length (i.e.~the inverse of its aspect ratio).  Shortly thereafter, Lighhill proposed a derivation  which was accurate to order $\epsilon^{1/2}$ \cite{1976,Childress1981}. This was a vast improvement on the earlier methods but still did not account for the ends of the body.  Eventually, Johnson derived a version of SBT that took into account the ends of the body \cite{Johnson1979}. He further showed that the force distribution obtained in his equations was accurate to order $\epsilon^{2}\log \epsilon$, which was achieved   by placing higher-order singularities along the centerline and matching the boundary conditions to higher order. These additional singularities  added no additional force to the system thus leaving the leading-order force distribution   unchanged. 

The SBT equations derived by Johnson for a slender filament of length $2\ell$ gives the velocity $\mathbf{U}(s)$ at a specific arc length $s$ along the filament ($-\ell \leq s \leq \ell$) as an integral
\begin{eqnarray} 
8 \pi \mu \mathbf{U}(s) &=&  \int_{-\ell}^{\ell} \left[ \frac{\mathbf{I}+ \mathbf{\hat{R}}_{0} \mathbf{\hat{R}}_{0}}{|\mathbf{R}_{0}|}\cdot \mathbf{f}(s') -\frac{\mathbf{I}+ \mathbf{\hat{t}\hat{t}}}{|s'-s|} \cdot \mathbf{f}(s) \right] \,d s' \notag \\ 
&&+\log\left(\frac{\ell^{2}(1-s^{2})}{r_{b}^{2} \rho(s)^{2} e}\right)\left(\mathbf{I}+ \mathbf{\hat{t}  \hat{t}}\right) \cdot \mathbf{f}(s)+2\left(\mathbf{I}- \mathbf{\hat{t}  \hat{t}}\right) \cdot \mathbf{f}(s),\label{SBT}
\end{eqnarray}
where  $e$ is the exponential, $2r_{b}$ is the thickness of the body, 
$\rho(s)$ is the dimensionless radial surface distribution (so that the surface of the body is located at $r = r_b \rho(s)$),  $\mathbf{R}_{0} = \mathbf{r}(s)-\mathbf{r}(s')$ is the vector between $s$ and $s'$ on the centerline, $\mathbf{\hat{t}}$ is the tangent to the centreline at $s$, and $\mathbf{f}$ the unknown force density along the centerline of the body.  This framework has been very successful in addressing many aspects of micro-scale fluid mechanics, in particular for the flow of fibers and self-propelled swimmers 
\cite{Koens2014,Tornberg2006,Kim2004,Lauga2006,Tornberg2004,Spagnolie2010,Spagnolie2011}.

In the current  paper, we perform an expansion similar to that of Johnson's SBT but for ribbons characterized by three length scales such that $\ell \gg b \gg a$. In order to do so, we place a plane of singularities within the ribbon, expand the system into the relevant regions, and asymptotically match them. As we demonstrate, the force distribution can be approximated with errors at most of  order $b/\ell$ without the need for any other singularity than the stokeslet. 

\section{Slender-ribbon theory} \label{sec:SREderivation}

\subsection{The slender ribbon geometry} \label{sec:ribbondescription}

\begin{figure} 
\begin{center}
\includegraphics[width=0.8\textwidth, bb= 0 0 964 624]{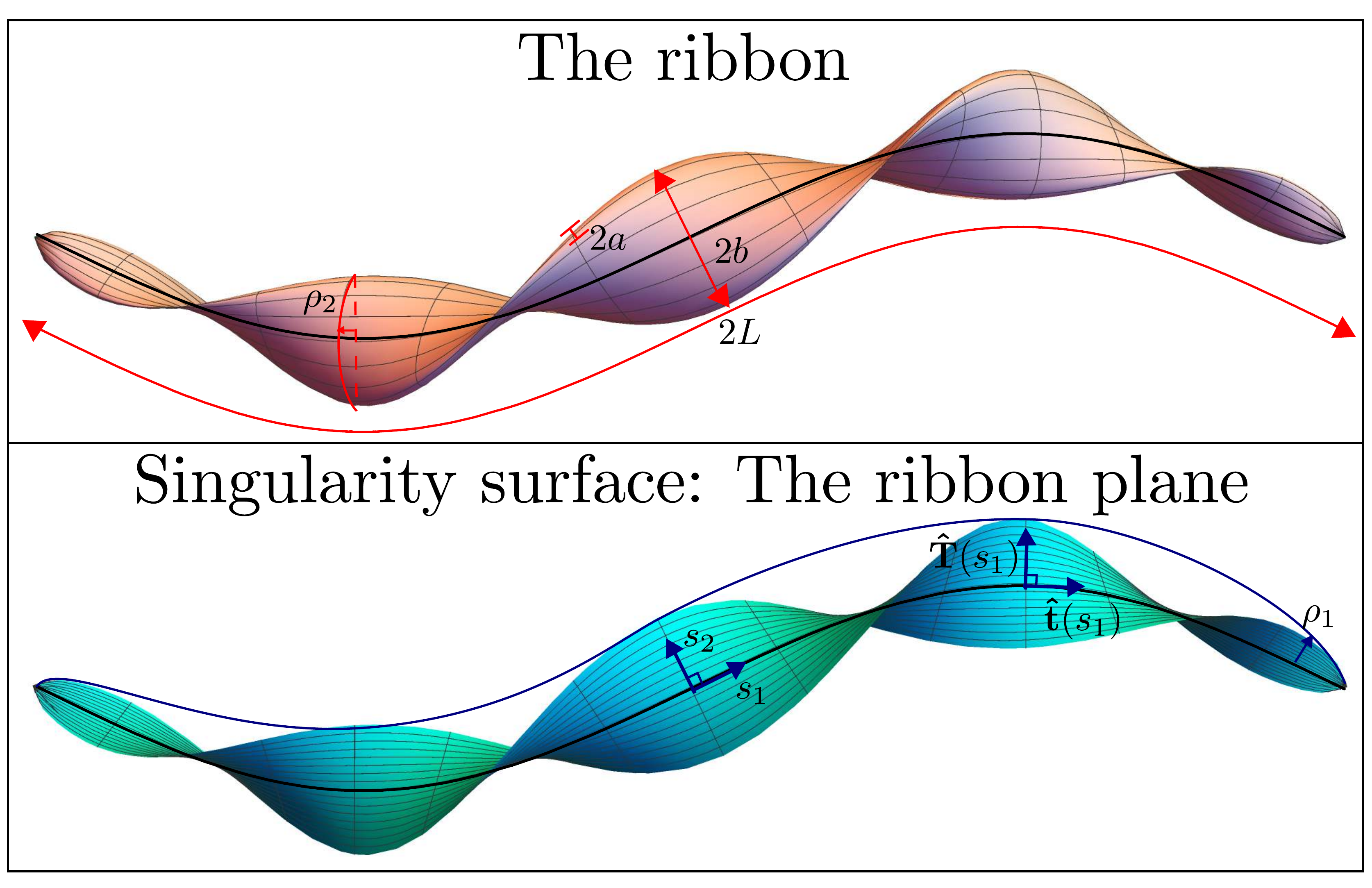}
\caption{Schematic representation of the full ribbon surface (top) and the equivalent ribbon plane (bottom). The black line in both figures depicts the centerline, described by points located at $\mathbf{r}(s_{1})$. The centerline length is  $2\ell$, the maximum length along the long edge (width) is $2b$, and the maximum length along the short edge (thickness) is $2a$. The functions $\rho_{1}$ and $\rho_{2}$ characterize how the long and short edge lengths vary across the ribbon, respectively. 
The vector $\mathbf{\hat{t}}(s_{1})$ is the tangent vector to the centerline and $\mathbf{\hat{T}}(s_{1})$ is the direction of the long edge sits in. The ribbon shape is parametrized using $s_{1}$, which describes the arclength along the centerline, and $s_{2}$, which gives the displacement in the direction $\mathbf{\hat{T}}(s_{1})$ from the centerline.}
\label{fig:shape}
\end{center}
\end{figure}

The ribbon structures considered in this paper have a centerline length of $2\ell$, a long edge (plane width) of length $2b$ and a short edge (thickness) of length $2a$ (see illustration in  Fig.~\ref{fig:shape}, top). The position of the centerline is given by the vector $\mathbf{r}(s_{1})$, where $s_{1}$ is the arclength along the centerline. At a given value of $s_{1}$, the long edge of the ribbon points in the direction $\mathbf{\hat{T}}(s_{1})$ and has an edge-to-edge width of $2 b \rho_{1}(s_{1})$. Here $\mathbf{\hat{T}}(s_{1})$ is a unit vector and $\rho_{1}(s_{1})$ contains the information of how the width varies along the ribbon length.  In our calculation, we assume that $\rho_{1}(s_{1})$ behaves like an ellipsoid near the ends of the ribbon and that $\mathbf{\hat{T}}(s_{1})$ remains everywhere perpendicular to the tangent of the centerline, $\mathbf{\hat{t}}(s_{1})$. Both are also assumed to  change smoothly with $s_{1}$.  The displacement from the centerline along the direction  $\mathbf{\hat{T}}(s_{1})$ is measured by a second arc length denoted $s_{2}$ ($ -b \rho_{1}(s_{1}) \leq s_{2} \leq b \rho_{1}(s_{1})$ ). Since $\mathbf{\hat{T}}(s_{1})$ does not depend on $s_{2}$ the long edge sits in a plane defined by $\mathbf{\hat{T}}(s_{1})$ and $\mathbf{\hat{t}}(s_{1})$. Physically this assumption means that the body is suitably rigid to prevent the bending along $s_{2}$.

The shape defined by the centerline and the vector  $\mathbf{\hat{T}}$  is called the ribbon plane. Any point on this plane is thus located at position $\mathbf{X}(s_{1},s_{2}) $ given by
\begin{equation} \label{scaledsheet}
\mathbf{X}(s_{1},s_{2}) = \mathbf{r}(s_{1}) + s_{2} \mathbf{\hat{T}}(s_{1}),
\end{equation}
as is illustrated in Fig.~\ref{fig:shape} (bottom).  The definition of this ribbon plane is consistent with the typical mathematical definition of a strip or ribbon \cite{Fuller1971}.
 Finally the surface of the physical ribbon itself is located a distance $\pm a \rho_{2} (s_{1},s_{2})$ normal to the ribbon plane at $s_{1}$, $s_{2}$. Again $\rho_{2}(s_{1},s_{2})$  is assumed to be roughly ellipsoidal towards the edges but also must be greater than 0 away from the edges. The surface of the ribbon is then described by material points $\mathbf{S}(s_{1},s_{2})$ given by
 \begin{equation}
\mathbf{S}(s_{1},s_{2}) = \mathbf{r}(s_{1}) + s_{2}   \mathbf{\hat{T}}(s_{1}) \pm a \rho_{2} (s_{1},s_{2}) \mathbf{\hat{n}},
\end{equation}
where $\mathbf{\hat{n}}$ is the normal to the ribbon plane (see notation in  Fig.~\ref{fig:shape}). Though this mathematical description can be used to describe ribbons with arbitrary thickness, the framework derived below for SRT only characterizes the fluid dynamic forces when the length is much larger than the width, which itself is much larger than the thickness, i.e.~the limit $\ell\gg b \gg a$. In this regime, surfaces can accurately be described by a ruled surface, as above. 

\subsection{An outline of the expansion} \label{sec:outline}
In this section we derive the leading order flow for bodies with slender-ribbon shapes. Like the expansion for SBT the total flow will be represented by a series of fundamental singularities and then expanded in the small parameters. Unlike SBT, singularities will here be placed within the ribbon plane, not just along the its centerline.

For slender-ribbon theory the singularities are placed in the ribbon plane for two reasons: the similarity between the cross-section of the ribbon and a prolate ellipsoids, and the requirement that the singularities used in a representation by fundamental singularities method must lie inside the body of the ribbon. Locally the ribbon is roughly cylindrical with a elliptical cross-section. This cross-section is similar to a central cross section of a prolate ellipsoid which has a flow given by a line of singularities placed between the two foci of the ellipsoid \cite{Chwang2006}. Therefore it is reasonable to assume that, that at least locally, the singularity distribution should be distributed over the ribbon plane. The requirement that the singularities must lie inside the body of the ribbon \cite{Kim2005} also supports this. As slender-ribbon theory expands the dynamics of the ribbon in the limit $\ell\gg b \gg a$, the zeroth order shape of the ribbon is the ribbon plane itself. Therefore, at zeroth order, the singularities can only be placed in the ribbon plane. 
 
Using the above justifications, the total fluid velocity at location $(s_{1},s_{2})$ on the surface of the ribbon arising from the hydrodynamic singularities in the plane is given by
\begin{equation}\label{eq9}
8 \pi \mu \mathbf{U}(s_{1},s_{2}) = 8 \pi \mu \int_{-\ell}^{\ell} \,d t_{1} \int_{-b \rho_{1}(t_{1})}^{b \rho_{1}(t_{1})} \,dt_{2} \left\lbrace \mathbf{U^{s}}(\mathbf{R};\mathbf{f}(t_{1},t_{2})) + \mathbf{U^{sd}}(\mathbf{R};\mathbf{g}_{SD}(t_{1},t_{2})) \right\rbrace,
\end{equation}
where $\mathbf{R} = \mathbf{X}(s_{1},s_{2}) -  \mathbf{X}(t_{1},t_{2}) \pm a \rho_{2}(s_{1},s_{2}) \mathbf{\hat{n}}(s_{1},s_{2})$, and the  integral  is naturally  broken into two parts: the component due to  the stokeslets, $\mathbf{U_{s}}$,  and the one due to the source dipoles, $\mathbf{U_{SD}}$. The above equation truncates the sum of singularities to stokeslets and source dipoles as this is all that is required for the leading order flow. At higher order it is likely that other singularities will be needed.

In order for the small parameters to be apparent in the formulation, we proceed to scale the integrals. 
The velocity is taken to scale as a typical velocity  $U$, the parameter $s_{2}$ as $b \rho_{1}$, and all other lengths as $\ell$. The total force on the ribbon will then scale as $\mu \ell U$ and the area of the sheet like $\ell b$, therefore the force per unit area,  $\mathbf{f}$, should scale as $\mu U/b $. 

The scaling of the source dipoles, $\mathbf{g}_{SD}$, in Johnson's (and Lighthill's) SBT derivation is proportional to the thickness of the body squared times the force per unit length. Therefore it is reasonable to  assume that the $\mathbf{g}_{SD}$ scaling would be a function of $b$ and $a$ with units length squared.
With this consideration, we then scale $\mathbf{g}_{SD}$ by $\mu U b \rho_{1}^{2}$. The scaled stokeslet and source dipole integrals are then given by
\begin{eqnarray}
8 \pi \mathbf{U_{s} }(s_{1},s_{2}) &=&  \int_{-1}^{1} \,d t_{1} \rho_{1}(t_{1}) \int_{-1}^{1} \,d t_{2} \left(\frac{ \mathbf{f}(t_{1},t_{2})}{|\mathbf{R}|} +\frac{ \mathbf{R} \mathbf{R}\cdot \mathbf{f}(t_{1},t_{2})}{|\mathbf{R}|^{3}}\right), \label{stokes}\\ 
8 \pi \mathbf{U_{SD} }(s_{1},s_{2}) &=& - b_{\ell}^{2}  \int_{-1}^{1} \,d t_{1} \rho_{1}(t_{1})^{3} \int_{-1}^{1}  \,d t_{2} \left( \frac{\mathbf{g}_{SD}(t_{1},t_{2})}{|\mathbf{R}|^{3}} -\frac{ 3 \mathbf{R} \mathbf{R} \cdot \mathbf{g}_{SD}(t_{1},t_{2})}{|\mathbf{R}|^{5}}\right), \label{source}
\end{eqnarray}
where
\begin{equation}
\mathbf{R}(s_{1},s_{2}) = \mathbf{R_{0}} + b_{\ell} \rho_{1}(s_{1}) s_{2} \mathbf{\hat{T}}(s_{1})  - b_{\ell} \rho_{1}(t_{1}) t_{2} \mathbf{\hat{T}}(t_{1}) \pm a_{\ell} \rho_{2}(s_{1},s_{2}) \mathbf{\hat{n}} (s_{1},s_{2}), \label{eq:R}
\end{equation}
and where we have denoted $b_{\ell} \equiv b/\ell$,  $a_{\ell}\equiv a/\ell$ and  $\mathbf{R_{0}} \equiv  \mathbf{r}(s_{1}) - \mathbf{r}(t_{1})$. The total force and torque are thus given by
  \begin{eqnarray}
  \mathbf{F} &=& \int_{-1}^{1} \,dt_{1}\int_{-1}^{1} \,dt_{2} \rho_{1}(t_{1}) \mathbf{f}(t_{1},t_{2}),\\
  \mathbf{L} &=&  \int_{-1}^{1} \,dt_{1}\int_{-1}^{1} \,dt_{2} \rho_{1}(t_{1}) \left[\mathbf{X}(t_{1},t_{2}) \times \mathbf{f}(t_{1},t_{2})\right],
  \end{eqnarray}
where the force $\mathbf{F}$ has been scaled by $ \mu \ell U$  and the torque $\mathbf{L}$  by $ \mu \ell^{2} U$.

A slender-body-like expansion requires the singularity integrals to be expanded in the relevant small parameters and then asymptotically-matched. In the case of a ribbon,  two small parameters are present, namely $b/\ell = b_{\ell}$ and $a/\ell = a_{\ell}$. The behavior of the hydrodynamic kernels in $s_{1}$ and $s_{2}$ should therefore each be expanded in three possible regions: $t-s = O(1)$, $t-s = O(b_{\ell})$, and $t-s =O(a_{\ell})$. The expansion procedure is outlined in Sec.~\ref{sec:expandpoints}, in which we show that only a subset of all expansions are actually required. 

After the solution has been expanded and asymptotically matched  (done in Sec.~\ref{sec: expansion}) the appropriate boundary conditions must be applied.  For a slender ribbon, we assume that material points on the surface immediately below and above a point in the ribbon plane (in the $\mathbf{\hat{n}}$ direction) move with the same velocity as that point. Physically this states that the ribbon surface does not expand, contract or shear. Mathematically this is  equivalent to ensuring that the final equations have no $\pm$ signs. The no-slip boundary condition for the fluid is applied and thus the velocity components of the material points on the surface are equal the velocity in the fluid there.

\subsection{Relevant expansion points} \label{sec:expandpoints}

\begin{figure}
\begin{center}
\includegraphics[width=0.9\textwidth, bb= 0 0 992 283]{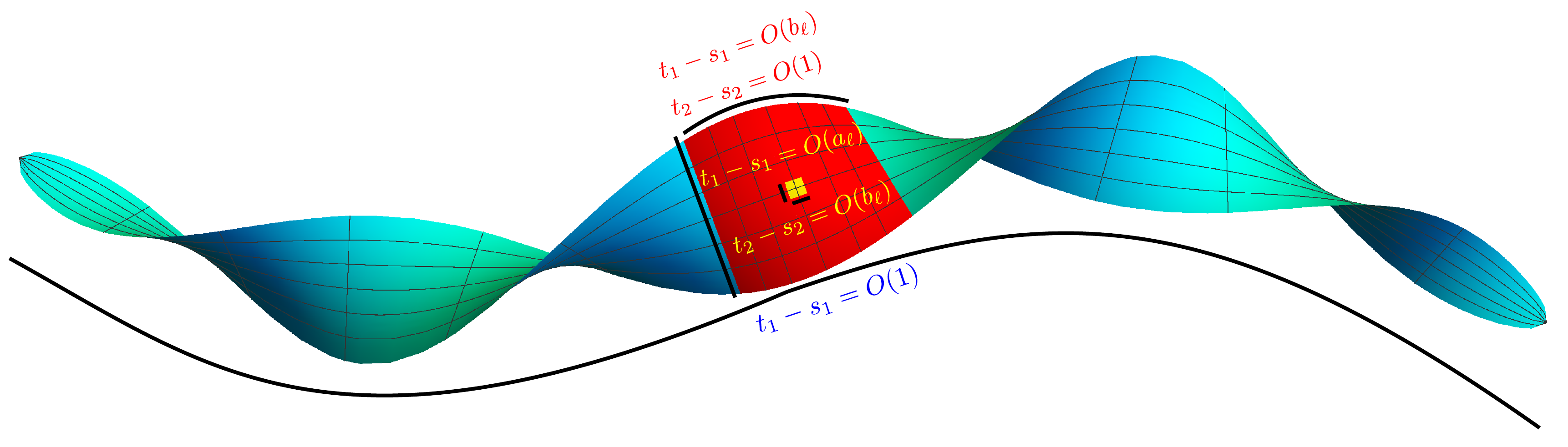}
\caption{A diagram depicting the scales of the different expansion regions measured about the  ribbon center. Most of the length is occupied by the outer or $O(1)$, $O(1)$ region in blue, dealing with interactions on that scale. The  red box depicts  the middle $O(1)$, $O(b_{\ell})$  region, which deals with variation on the scale of the square. Finally the  small yellow box represents the inner  $O(b_{\ell})$, $O(a_{\ell})$ region.}
\label{fig:regions}
\end{center}
\end{figure}

In Johnson's SBT the singularity kernels are integrated over one dimension, namely the arclength $s_{1}$. Depending on how $s_{1}$ scales, the behavior of the singularity kernels are distinct. Specifically, when  $s_{1}$ scales with the total arclength along the slender body, the influence of singularities far from the point of interest contribute; in contrast,  if $s_{1}$ scales with the body thickness, then  the local contribution from the singularities are important. By combing these two regions and removing any common behavior found in the overlap region (the common part), the full integrand can be evaluated asymptotically. 

These two different regions reflect the two length scales inherent to the slender body formulation.  In the slender-ribbon formulation the kernels are integrated over two dimensions, $s_{1}$ and $s_{2}$, and have three (dimensionless) length scales, 1, $b_{\ell}$ and $a_{\ell}$. Therefore there are nine possible forms the kernel can take. However, many of these nine kernels are simply a limiting case of another one of the nine. In that case, the common asymptotic behavior of these limit regions will be identical to their behavior outside the overlap region. Therefore these kernels are redundant and will be removed by the subtraction of their common parts. Performing the calculation on each region and then relying on the removal of the redundant parts by matching is very mathematically intensive. If the  regions relevant to the final solution can be determined beforehand, such a long derivation is not needed. 

The scaled stokeslet and source dipole kernels only depend on the value of $\mathbf{R}$, $\mathbf{f}$ and $\mathbf{g}_{SD}$. A Taylor series expansion of $\mathbf{f}$ and $\mathbf{g}_{SD}$ give their respective forms in each of the nine regions, so their variation occurs in a very obvious manner. The leading-order behavior of $\mathbf{R}$ can change significantly between the different regions and therefore is a good indicator of when the behavior of the kernels change. In Table~\ref{tab:R} we show the leading-order form of $\mathbf{R}$ in the different limits.

\begin{table}[t]
\begin{tabular}{|c|c c c|} \hline
$\mathbf{R}$& $t_{2} -s_{2} =a_{\ell} \chi_{2,a_{\ell}}$ & $t_{2} -s_{2} =b_{\ell} \chi_{2,b_{\ell}}$ & $t_{2} -s_{2} =O(1)$ \\ \hline
$t_{1} -s_{1} =O(1)$  & $ \mathbf{R_{0}} $  & $ \mathbf{R_{0}} $ & \cellcolor{cyan!25}$ \mathbf{R_{0}} $ \\   
$t_{1} -s_{1} =b_{\ell} \chi_{1,b_{\ell}}$ & $  - b_{\ell} \chi_{1,b_{\ell}} \mathbf{\hat{t}}$ &  $  - b_{\ell} \chi_{1,b_{\ell}} \mathbf{\hat{t}}$ &\cellcolor{red!25} $  b_{\ell} \left(-\chi_{1,b_{\ell}} \mathbf{\hat{t}} + \rho_{1}(s_{1}) \mathbf{\hat{T}}(s_{1})(s_{2} - t_{2})\right)$  \\  
 \multirow{2}{*}{$t_{1} -s_{1} =a_{\ell} \chi_{1,a_{\ell}}$} & \multirow{2}{*}{ $-a_{\ell} \chi_{1,a_{\ell}} \mathbf{\hat{t}}\pm a_{\ell} \rho_{2} \mathbf{\hat{n}}$ } & \cellcolor{yellow!25} $-a_{\ell} \chi_{1,a_{\ell}} \mathbf{\hat{t}}\pm a_{\ell} \rho_{2} \mathbf{\hat{n}}$  &  \multirow{2}{*}{ $  b_{\ell} \rho_{1}(s_{1}) \mathbf{\hat{T}}(s_{1})(s_{2} - t_{2})$ }\\
 & & \cellcolor{yellow!25}$ + b_{\ell}^{2} \rho_{1}(s_{1}) \chi_{2,b_{\ell}} \mathbf{\hat{T}}(s_{1})$  & \\ \hline
\end{tabular}
\caption{The  expansion of $\mathbf{R}$ in the nine different asymptotic regions. When the $t-s$ values are of order $b_{\ell}$ or $a_{\ell}$ the expressions are obtained  by  Taylor series expansions of the relevant terms. Each colored block represents a region in which the kernel must be expanded. Blue represents the  $t_2-s_2 = O(1)$, $t_1-s_1 = O(1)$ (outer) region, red represents the $O(1)$, $O(b_{\ell})$ (middle) region, and yellow represents the $O(b_{\ell})$ $O(a_{\ell})$ (inner) region.}
\label{tab:R}
\end{table}
 
The top row of Table~\ref{tab:R} shows that in all these regions $\mathbf{R}$ takes the same form at leading order. Only one of these three regions is needed as the common behavior between them would remove the redundant parts. The top row right column (namely the $t_2-s_2 = O(1)$, $t_1-s_1 =O(1)$ region) is the completely unexpanded region, and so $\mathbf{f}$ and $\mathbf{g}_{SD}$ take their most general form there. Hence the top row, left and center columns have behavior which is already contained within the $O(1)$, $O(1)$ region. These redundant regions obviously would not add anything new to the expansion. Therefore, only the $O(1)$, $O(1)$ region needs to be included from the top-row. This is the region highlighted in blue in the table.

A similar procedure is needed for the remaining six regions, however the common behavior there is not as obvious. For example the form of $\mathbf{R}$ in the right column of the center row (namely the $O(1)$, $O(b_{\ell})$ region) limits to the form in the left and center columns. Therefore, though the form is not exactly the same, this row can be correctly represented by the $O(1)$, $O(b_{\ell})$ region. This is the region highlighted in red in the table.

 Finally in the bottom row, the $\mathbf{R}$ of the center column (namely the $O(b_{\ell})$, $O(a_{\ell})$ region) is seen to correctly limit to the other regions in this row. However in this region $\mathbf{f}$ and $\mathbf{g}_{SD}$ are expanded with respect to $s_{2}$ in factors of $b_{\ell}$. This expansion prevents the center column kernel representing the far right column. Conveniently, the $\mathbf{R}$ of the right column of the bottom row is a limit of the $O(1)$, $O(b_{\ell})$ region. Therefore the right column of the bottom row does not need to be accounted for, as it already is accounted for in the $O(1)$, $O(b_{\ell})$ region. Ignoring the bottom row right column term, the $O(b_{\ell})$ $O(a_{\ell})$ region term correctly captures the remaining behavior of the other terms in this row (yellow).

The above analysis suggests that there are three distinct asymptotic regions for $t_2-s_2$ and $t_1-s_1$ to expand the kernels in: the $O(1)$, $O(1)$ region, the $O(1)$, $O(b_{\ell})$ region, and the $O(b_{\ell})$, $O(a_{\ell})$ region. For simplicity these regions will be called the outer, middle and inner regions respectively, and they are illustrated schematically in Fig.~\ref{fig:regions}. 
 The stokeslet kernel will now be expanded in these three regions and the common parts subtracted, followed by the final integration. 
Inspecting the final result we will see that the source dipole distribution will in fact  not be needed for the leading-order flow.

\subsection{The leading-order expansion} \label{sec: expansion}

The leading-order hydrodynamic behavior of a slender ribbon can be found by expanding the stokeslet kernel in the outer, middle and inner regions, and  removing the common behavior found in the overlap regions. This is then followed by the integration of the asymptotic kernels and the use of the boundary conditions to obtain the final asymptotic approximation.

From Eq.~\eqref{stokes} the stokeslet kernel has the form
\begin{equation}
\mathbf{K}_{S} =\rho_{1}(t_{1})\left( \frac{ \mathbf{f}(t_{1},t_{2})}{|\mathbf{R}|} +\frac{ \mathbf{R} \mathbf{R}\cdot \mathbf{f}(t_{1},t_{2})}{|\mathbf{R}|^{3}}\right),
\end{equation} 
where $\mathbf{R}$ is defined by Eq.~\eqref{eq:R}. We proceed to expand this term in the three regions and remove the overlap.

\subsubsection{Outer region} \label{sec:sout}

In the outer,  $O(1)$, $O(1)$ region, $ t_{1} -s_{1} = O(1)$ and $ t_{2} -s_{2} = O(1)$.  In this region $\mathbf{R}$ is approximately given by
\begin{equation}
\mathbf{R}^{(o)} = \mathbf{R_{0}}  + b_{\ell}(s_{2} \rho_{1}(s_{1}) \mathbf{\hat{T}}(s_{1}) - t_{2} \rho_{1}(t_{1}) \mathbf{\hat{T}}(t_{1})) + O(b_{\ell}^{2}) + O(a_{\ell}),
\end{equation}
where the superscript $(o)$ indicates an expansion in terms of the outer region.
We thus have 
\begin{eqnarray}
|\mathbf{R}|^{2} &=& |\mathbf{R}_{0}|^{2} + 2 b_{\ell} \mathbf{R_{0}} \cdot (s_{2} \rho_{1}(s_{1}) \mathbf{\hat{T}}(s_{1}) - t_{2} \rho_{1}(t_{1}) \mathbf{\hat{T}}(t_{1})) + O(b_{\ell}^{2}) + O(a_{\ell}), \\
|\mathbf{R}|^{-1} &=& \frac{1}{|\mathbf{R}_{0}|} - \frac{b_{\ell} \mathbf{R_{0}} \cdot (s_{2} \rho_{1}(s_{1}) \mathbf{\hat{T}}(s_{1}) - t_{2} \rho_{1}(t_{1}) \mathbf{\hat{T}}(t_{1}))}{|\mathbf{R}_{0}|^{3}} + O(b_{\ell}^{2}) + O(a_{\ell}), \\
|\mathbf{R}|^{-3} &=& \frac{1}{|\mathbf{R}_{0}|^{3}} - 3 \frac{b_{\ell} \mathbf{R_{0}} \cdot (s_{2} \rho_{1}(s_{1}) \mathbf{\hat{T}}(s_{1}) - t_{2} \rho_{1}(t_{1}) \mathbf{\hat{T}}(t_{1}))}{|\mathbf{R}_{0}|^{5}} + O(b_{\ell}^{2}) + O(a_{\ell}), \\
\mathbf{R} \mathbf{R} &=&  \mathbf{R_{0}}\mathbf{R_{0}} +  b_{\ell} \left[\mathbf{R_{0}} (s_{2} \rho_{1}(s_{1}) \mathbf{\hat{T}}(s_{1}) - t_{2} \rho_{1}(t_{1}) \mathbf{\hat{T}}(t_{1})) + (s_{2} \rho_{1}(s_{1}) \mathbf{\hat{T}}(s_{1}) - t_{2} \rho_{1}(t_{1}) \mathbf{\hat{T}}(t_{1}))\mathbf{R_{0}} \right] \nonumber \\ && +O(b_{\ell}^{2}) + O(a_{\ell}),
\end{eqnarray}
and the leading-order outer-region stokeslet kernel is given by
\begin{equation}
\mathbf{K}_{S}^{(o)} = \rho_{1}(t_{1})\left( \frac{ \mathbf{f}(t_{1},t_{2})}{|\mathbf{R}_{0}|} +\frac{ \mathbf{R_{0}} \mathbf{R_{0}}\cdot \mathbf{f}(t_{1},t_{2})}{|\mathbf{R}_{0}|^{3}}\right)  +O(b_{\ell}) + O(a_{\ell}). 
\end{equation}

\subsubsection{Middle region} \label{sec:smid}

In the middle, $O(1)$, $O(b_{\ell})$ region,  $t_{1}-s_{1}=O(b_{\ell})$. In this region, any dependence on $t_{1}$ should be written in terms of the behavior at $s_{1}$. This is done by using a Taylor series expansion around $s_{1}$, for all the terms which depend on $t_{1}$. We define the new scaled variable 
\begin{equation}
\chi_{1,b_{\ell}} = \frac{t_{1} -s_{1}}{ b_{\ell}}\cdot
\end{equation}
The terms in the stokeslet kernel which depend on $t_{1}$ are $\mathbf{r}(t_{1})$, $\rho_{1}(t_{1})$, $\mathbf{\hat{T}}(t_{1})$, and $\mathbf{f}(t_{1},t_{2})$. Their Taylor series expansions are given by
\begin{eqnarray}
\mathbf{r}(t_{1}) &=& \mathbf{r}(s_{1}) + (t_{1}-s_{1}) \mathbf{\hat{t}} + \frac{(t_{1}-s_{1})^{2} \kappa}{2} \mathbf{\hat{n}_{s_{1}}} + ..., \\
&=& \mathbf{r}(s_{1}) + b_{\ell} \chi_{1,b_{\ell}}  \mathbf{\hat{t}} + b_{\ell}^{2} \chi_{1,b_{\ell}}^{2} \frac{\kappa}{2} \mathbf{\hat{n}_{s_{1}}} + O(b_{\ell}^{3}), \\
\mathbf{R_{0}} &=& -  b_{\ell} \chi_{1,b_{\ell}}  \mathbf{\hat{t}} - b_{\ell}^{2} \chi_{1,b_{\ell}}^{2} \frac{\kappa}{2} \mathbf{\hat{n}_{s_{1}}} + O(b_{\ell}^{3}),\\ 
\rho_{1}(t_{1}) &=& \rho_{1}(s_{1}) +   b_{\ell} \chi_{1,b_{\ell}} \partial_{s_{1}} \rho_{1}(s_{1}) + O(b_{\ell}^{2}), \\
\mathbf{\hat{T}}(t_{1}) &=& \mathbf{\hat{T}}(s_{1}) + b_{\ell} \chi_{1,b_{\ell}} \sigma \mathbf{\hat{N}} + O(b_{\ell}^{2}), \\
\mathbf{f}(t_{1},t_{2}) &=& \mathbf{f}(s_{1},t_{2}) + b_{\ell} \chi_{1,b_{\ell}}   \partial_{s_{1}}\mathbf{f}(s_{1},t_{2}) + O(b_{\ell}^{2}), 
\end{eqnarray}
where $\mathbf{\hat{n}_{s_{1}}}$ is the normal vector to the centerline, $\kappa$ is the curvature of the centerline, $\mathbf{\hat{N}}$ is the direction of twisting of the ribbon plane and $\sigma \mathbf{\hat{N}} = \partial_{s_{1}} \mathbf{\hat{T}}$.
From the above series, $\mathbf{R}$ is
\begin{multline}
\frac{\mathbf{R}^{(m)}}{b_{\ell}} = -\chi_{1,b_{\ell}} \mathbf{\hat{t}} + \rho_{1}(s_{1}) \mathbf{\hat{T}}(s_{1})(s_{2} - t_{2}) \\
+b_{\ell} \left( -  \chi_{1,b_{\ell}}^{2} \frac{\kappa}{2} \mathbf{\hat{n}_{s_{1}}} - \sigma \rho_{1}(s_{1}) t_{2} \chi_{1,b_{\ell}} \mathbf{\hat{N}} - t_{2} \chi_{1,b_{\ell}}  (\partial_{s_{1}}\rho_{1}(s_{1})) \mathbf{\hat{T}}(s_{1}) \pm \frac{a_{\ell}}{b_{\ell}^{2}} \rho_{2} \mathbf{\hat{n}} \right)\\ + O(b_{\ell}^{2}) + O(a_{\ell}),
\end{multline}
while the leading-order kernel is given by
\begin{equation}
b_{\ell} \mathbf{K}_{S}^{(m)} =  \rho_{1}(s_{1})\left( \frac{ \mathbf{I}}{|\mathbf{R}^{(m)}_{1}|} +\frac{ \mathbf{R}^{(m)}_{1} \mathbf{R}^{(m)}_{1}}{|\mathbf{R}^{(m)}_{1}|^{3}}\right) \cdot \mathbf{f}(s_{1},t_{2}) + O(b_{\ell}) + O(a_{\ell}),
\end{equation}
with
\begin{eqnarray}
\mathbf{R}^{(m)}_{1} &=&  -\chi_{1,b_{\ell}} \mathbf{\hat{t}} + \rho_{1}(s_{1}) \mathbf{\hat{T}}(s_{1})(s_{2} - t_{2}),
\end{eqnarray}
and with the superscript $(m)$ to indicate the middle expansion. 

Importantly, in this middle expansion, terms proportional to the curvature and rate of twisting terms have been discarded. These terms scale like $b_{\ell} \kappa$ and $b_{\ell} \sigma$, respectively. Therefore slender-ribbon theory assumes that the curvature, $\kappa$, and rate of twisting, $\sigma$, are less than O($b_{\ell}^{-1}$). If $\kappa$ or $\sigma$ do become large the error associated with these terms could also become large. The condition on $\sigma$ is not hard to satisfy, as ribbon with $b_{\ell}=10^{-2}$ would require around 30 twists to have a mean $\sigma$ of O($b_{\ell}^{-1}$). However, like in SBT, it is easier to think of configurations where $\kappa$ becomes large. The centerline bending condition here is identical to that required in the framework of SBT. We note that SBT has been used very successfully in applications where this curvature condition has been broken \cite{Koens2014}. Therefore slender-ribbon theory could also work when $\kappa$ or $\sigma$ are large but the results should be viewed with caution as they are formally outside the expected domain of validity.

\subsubsection{Inner region} \label{sec:sin}

The last region to expand is the inner or $O(b_{\ell})$, $O(a_{\ell})$ region.  In this region, terms with $t_{1}$ or $t_{2}$ must be represented by  series. Unlike the middle region $t_{1}$ is expanded in powers of $a_{\ell}$ while $t_{2}$ is expanded in powers of $b_{\ell}$. Similarly to the middle region, two new variables are defined,
\begin{eqnarray}
\chi_{1,a_{\ell}} = \frac{t_{1} -s_{1}}{a_{\ell}},  \\
\chi_{2,b_{\ell}} = \frac{t_{2} - s_{2}}{b_{\ell}}\cdot
\end{eqnarray}
Using these variables and recognizing that the expanded terms in the middle region have a similar form in the inner part, $\mathbf{R}$ can be shown to be given by
\begin{equation}
\frac{\mathbf{R}^{(i)}}{a_{\ell}} = -\chi_{1,a_{\ell}} \mathbf{\hat{t}} \pm \rho_{2} \mathbf{\hat{n}}  - \frac{b_{\ell}^{2}}{a_{\ell}} \chi_{2,b_{\ell}} \rho_{1} \mathbf{\hat{T}}
- b_{\ell} s_{2} \chi_{1,a_{\ell}} \left( \mathbf{\hat{T}} \partial_{s_{1}} \rho_{1} + \rho_{1} \sigma \mathbf{\hat{N}} \right) + O(b_{\ell}^{2}) + O(a_{\ell}).
\end{equation}
Hence the stokeslet kernel in the inner region is given by
\begin{equation}
a_{\ell} \mathbf{K}_{S}^{(i)} =\rho_{1}(s_{1})\left( \frac{ \mathbf{I}}{|\mathbf{R}^{(i)}_{1}|} +\frac{ \mathbf{R}^{(i)}_{1} \mathbf{R}^{(i)}_{1}}{|\mathbf{R}^{(i)}_{1}|^{3}}\right) \cdot  \mathbf{f}(s_{1},s_{2})  + O(b_{\ell}) + O(a_{\ell}),
\end{equation}
where 
\begin{eqnarray}
\mathbf{R}^{(i)}_{1} &=&  -\chi_{1,a_{\ell}} \mathbf{\hat{t}} \pm \rho_{2} \mathbf{\hat{n}}  - \frac{b_{\ell}^{2}}{a_{\ell}} \chi_{2,b_{\ell}} \rho_{1} \mathbf{\hat{T}},
\end{eqnarray}
and again we used the the superscript $(i)$ to indicate part of the inner expansion.

\subsubsection{Overlap regions}
The common behavior between the above kernels needs to be subtracted to get the complete asymptotic solution for the stokeslets. This is done, classically, by  expanding one kernel in terms of another region's variables and vice versa (done to check consistency).   
The overlap region that is most similar to that of SBT is the overlap between the outer (Sec.~\ref{sec:sout}) and middle regions (Sec.~\ref{sec:smid}).  The outer kernel expanded in terms of the middle variables looks like
\begin{equation}
\mathbf{K}_{S}^{(o) \in (m)} =  \frac{ \mathbf{I} +\mathbf{\hat{t}} \mathbf{\hat{t}} }{ b_{\ell} |\chi_{1,b_{\ell}}|}   \cdot \rho_{1}(s_{1})  \mathbf{f}(s_{1},t_{2})  +O(1) + O(a_{\ell} b_{\ell}^{-1} ),
\end{equation}
where we used the symbol ``$(o) \in (m)$'' to indicate that it is the outer kernel expanded in terms of the middle variables (with a labelling convention similar in other cases). The middle kernel expanded in terms of the outer variables looks like
\begin{equation} \label{commonom}
 \mathbf{K}_{S}^{(m) \in (o)} =  \rho_{1}(s_{1}) \frac{ \mathbf{I} + \mathbf{\hat{t}} \mathbf{\hat{t}}}{|t_{1} - s_{1}|} \cdot \mathbf{f}(s_{1},t_{2})+ O(b_{\ell}) + O(a_{\ell}).
\end{equation}
As expected, these two terms are identical. Furthermore, we note that  Eq.~\eqref{commonom} is almost identical to the expression for the  overlap region in SBT \cite{Johnson1979}, showing expected similarities in the derivations.

In contrast, the common behavior between the middle and inner region has no equivalent in SBT since the inner region reflects the third length scale of the problem (ribbon thickness). Using the expressions in Sec.~\ref{sec:sin}, we obtain the leading-order $\mathbf{R}$ behavior when expanding the middle kernel in terms of the inner variables as
\begin{eqnarray}
b_{\ell} \mathbf{R}_{1}^{(m) \in (i)} &=& a_{\ell} ( - \chi_{1,a_{\ell}} \mathbf{\hat{t}} - \frac{b_{\ell}^{2}}{a_{\ell}} \chi_{2,b_{\ell}} \rho_{1} \mathbf{\hat{T}}), \nonumber \\
&=& a_{\ell} \mathbf{R}_{a},
\end{eqnarray}
which gives a common kernel of
\begin{equation}
\mathbf{K}_{S}^{(m)\in (i)} = \frac{\rho_{1}}{a_{\ell}}\left( \frac{ \mathbf{I}}{| \mathbf{R}_{a}|} +\frac{ \mathbf{R}_{a} \mathbf{R}_{a}}{|\mathbf{R}_{a}|^{3}}\right) \cdot  \mathbf{f}(s_{1},s_{2})  + O(b_{\ell} a_{\ell}^{-1}) + O(a_{\ell}^{0}).
\end{equation}
Similarly, when expanding the inner kernel in terms of the middle variables we obtain
\begin{eqnarray}
a_{\ell} \mathbf{R}^{(i)}_{1} &=&  b_{\ell}(- \chi_{1,b_{\ell}} \mathbf{\hat{t}}  - (t_{2}-s_{2}) \rho_{1} \mathbf{\hat{T}}) \pm a_{\ell} \rho_{2} \mathbf{\hat{n}}, \nonumber \\
&=& b_{\ell} \mathbf{R}_{a}^{(m)} \pm a_{\ell} \rho_{2} \mathbf{\hat{n}} ,
\end{eqnarray}
which gives a common kernel of
\begin{equation}
 \mathbf{K}_{S}^{(i) \in (m)} =\rho_{1}\left(\frac{\mathbf{I}  }{ b_{\ell} |\mathbf{R}_{a}^{(m)} |}+\frac{ \mathbf{R}_{a}^{(m)} \mathbf{R}_{a}^{(m)} }{ b_{\ell} |\mathbf{R}_{a}^{(m)} |^{3}} \right) \cdot  \mathbf{f}(s_{1},s_{2})
+ O(1) + O(a_{\ell} b_{\ell}^{-1}).
\end{equation}
The two kernels are not as manifestly identical as in the outer and middle case, but by expanding them explicitly it is easily shown  that they are.

Note finally that the overlap of the outer and inner regions does not need to be considered since the  behavior  common to outer and inner regions is included within the overlap of the middle and the inner regions.

\subsubsection{Complete stokeslet kernel}

From the above expansions, the complete stokeslet kernel is then asymptotically approximated by
\begin{equation}
\mathbf{K}_{S} \approx \mathbf{K}_{S}^{(o)} +\mathbf{K}_{S}^{(m)} + \mathbf{K}_{S}^{(i)} - \mathbf{K}_{S}^{(m) \in (o)} - \mathbf{K}_{S}^{(m) \in (i)}, 
\end{equation}
which when expanded gives
\begin{eqnarray} \label{1storder}
\mathbf{K}_{S} &\approx &
\notag  \frac{ \mathbf{I} + \mathbf{\hat{R}_{0}} \mathbf{\hat{R}_{0}}}{|\mathbf{R}_{0}|} \cdot \rho_{1}(t_{1}) \mathbf{f}(t_{1},t_{2})   -  \frac{ \left(\mathbf{I} + \mathbf{\hat{t}} \mathbf{\hat{t}}\right)}{|t_{1}-s_{1}|} \cdot  \rho_{1}(s_{1}) \mathbf{f}(s_{1},t_{2})  \\ &&
 +   \frac{ \mathbf{I} + \mathbf{\hat{R}}^{(m)}_{1} \mathbf{\hat{R}}^{(m)}_{1}}{ b_{\ell} |\mathbf{R}^{(m)}_{1}|} \cdot  \rho_{1}(s_{1}) \mathbf{f}(s_{1},t_{2}) 
\notag + 
  \frac{ \mathbf{I}+  \mathbf{\hat{R}}^{(i)}_{1} \mathbf{\hat{R}}^{(i)}_{1}}{a_{\ell} |\mathbf{R}^{(i)}_{1}|} \cdot \rho_{1}(s_{1}) \mathbf{f}(s_{1},s_{2}) \\
&& - \left\{
 \frac{ \mathbf{I} +\mathbf{\hat{R}}_{a} \mathbf{\hat{R}}_{a}}{a_{\ell} | \mathbf{R}_{a}|}    \cdot  \rho_{1}(s_{1}) \mathbf{f}(s_{1},s_{2}) \right\}.
\end{eqnarray}

\subsection{The slender-ribbon equations} \label{sec:SRE}
By integrating Eq.~\eqref{1storder} over both $t_{1}$ and $t_{2}$ the asymptotic behavior of a sheet of stokeslets can then  be obtained.
The leading-order solution is thus given by the surface integral
\begin{eqnarray}
\notag 8 \pi \mathbf{U_{s} }(s_{1},s_{2}) = \int_{-1}^{1} \,d t_{1} \int_{-1}^{1} \,d t_{2}&&  \left[ \frac{ \mathbf{I} + \mathbf{\hat{R}_{0}} \mathbf{\hat{R}_{0}}}{|\mathbf{R}_{0}|} \cdot \rho_{1}(t_{1}) \mathbf{f}(t_{1},t_{2})   -  \frac{ \left(\mathbf{I} + \mathbf{\hat{t}} \mathbf{\hat{t}}\right)}{|t_{1}-s_{1}|} \cdot  \rho_{1}(s_{1}) \mathbf{f}(s_{1},t_{2})   \right.\\&& 
 +   \frac{ \mathbf{I} + \mathbf{\hat{R}}^{(m)}_{1} \mathbf{\hat{R}}^{(m)}_{1}}{ b_{\ell} |\mathbf{R}^{(m)}_{1}|} \cdot  \rho_{1}(s_{1}) f(s_{1},t_{2}) 
\notag \\ && + 
\notag   \frac{ \mathbf{I}+  \mathbf{\hat{R}}^{(i)}_{1} \mathbf{\hat{R}}^{(i)}_{1}}{a_{\ell} |\mathbf{R}^{(i)}_{1}|} \cdot \rho_{1}(s_{1}) \mathbf{f}(s_{1},s_{2}) 
\\&& 
\left. - 
 \frac{ \mathbf{I} +\mathbf{\hat{R}}_{a} \mathbf{\hat{R}}_{a}}{a_{\ell} | \mathbf{R}_{a}|}    \cdot  \rho_{1}(s_{1}) \mathbf{f}(s_{1},s_{2})   \right].
\end{eqnarray}

Wherever possible, we wish to evaluate the above integrals explicitly. However, this cannot be done for two of the terms.  
Specifically,  the first right-hand side term on the first line involves $f(t_{1},t_{2})$, and cannot be integrated over $t_{1}$ while  
second line involves $f(s_{1},t_{2})$ and therefore cannot be integrated over $t_{2}$. As a side remark, we note that each right-hand side  term on the first line is individually  singular  but when combined together the singularity disappears, a feature also appearing in SBT.

The remaining integrals, which can be evaluated, typically take the form
\begin{equation}
\int_{-1}^{1} \,dt_{1} \frac{\chi^{i}}{\sqrt{\chi^{2} + \theta^{2}}^{j}},
\end{equation}
where, $i$ and $j$ are positive integers, $\epsilon \chi = t_{1} -s_{1}$, $\theta$ is an arbitrary real function that does not depend on $t_{1}$, and $\epsilon$ is a small parameter. The asymptotic forms of these integrals have been evaluated previously by Gotz \cite{Gotz2000} and are listed in appendix~\ref{Appendix}.
After all possible integrations are performed, the final integral formulation for slender-ribbon theory is given by
\begin{eqnarray} \label{SREs}
\notag 8 \pi \mathbf{U }(s_{1},s_{2}) &=& \int_{-1}^{1} \,d t_{1}  \left[ \frac{ \mathbf{I} + \mathbf{\hat{R}_{0}} \mathbf{\hat{R}_{0}}}{|R_{0}|} \cdot \rho_{1}(t_{1}) \left\langle\mathbf{f}\right\rangle(t_{1})   -  \frac{ \left(\mathbf{I} + \mathbf{\hat{t}} \mathbf{\hat{t}}\right)}{|t_{1}-s_{1}|} \cdot  \rho_{1}(s_{1}) \left\langle\mathbf{f}\right\rangle(s_{1})  \right] \\&&
 +  \int_{-1}^{1} \,d t_{2} \log\left(\frac{4(1-s_{1}^{2})}{b_{\ell}^{2} \rho_{1}^{2} (s_{2}-t_{2})^{2}}\right)  \left(\mathbf{I} +\mathbf{\hat{t}} \mathbf{\hat{t}} \right) \notag \cdot  \rho_{1}(s_{1}) \mathbf{f}(s_{1},t_{2})\\ &&
 +2    \left(\mathbf{\hat{T}} \mathbf{\hat{T}}- \mathbf{\hat{t}} \mathbf{\hat{t}}\right) \cdot  \rho_{1}(s_{1}) \left\langle\mathbf{f}\right\rangle(s_{1}),
\end{eqnarray}
where we have used the notation $ \left\langle \mathbf{f} \right\rangle (t_{1})\equiv \int_{-1}^{1} \,d t_{2} \mathbf{f}(t_{1},t_{2}) $. This is the main result of our paper. Errors between this asymptotic result and the exact solution are at most of order $b_\ell$.

We note that in the final equation there is no $\pm$ signs and thus each point of the surface moves rigidly with a corresponding point on the ribbon plane. As this was the boundary condition we wished to enforce, no further singularities (source-dipoles) are needed in order to satisfy the boundary conditions. This is consistent with known results for the motion of rigid prolate spheroids in the small-thickness limit \cite{Chwang2006}.

 \subsection{Slender-ribbon versus slender body} \label{sec:SRTvsSBT}
  There are many similarities between our slender-ribbon equations (Eq.~\ref{SREs}) and Johnson's slender-body equations  (Eq.~\ref{SBT}). The left-hand sides of both equations have the exact same form and the integral over the centerline, $s_{1}$, is very similar. In fact, the integral over the centerline would be identical if Johnson's force distribution was replaced by $\rho_{1}(t_{1}) \left\langle\mathbf{f}\right\rangle(t_{1})$. Physically this means that in the far-field, the  ribbon behaves  like a slender body with a force density,  for each $s_{1}$, 
    equal to the total force across $s_{2}$. 
  
The logarithmic term in SRT is also very similar to the logarithmic term in SBT. Both these terms have the same tensorial behavior and contain a logarithm which depends on the aspect ratio and on how the surface of the body varies along its length.  However, unlike Johnson's SBT, the logarithm in the slender-ribbon equations has a dependence on the ribbon width, $s_{2}$. It also multiplies the force distribution and is integrated over. 
  
Finally, the remaining (non-integral) term in the SRT equations bears some resemblance to the remaining terms in Johnson's SBT. However instead of the tensor $\mathbf{\hat{T}} \mathbf{\hat{T}}$, Johnson's result has the identity tensor, $\mathbf{I}$. In  SBT, this local term modifies the logarithmic behaviour to further separate the drag for motion perpendicular and parallel to the centerline. 
  The same reasoning is at play here, however instead of just two directions (normal and tangential), three distinct directions must be considered. The $\mathbf{\hat{T}} \mathbf{\hat{T}}$ term therefore exists to ensure that the drag from motion in the normal $\mathbf{\hat{n}}$ direction is larger than that in the $\mathbf{\hat{T}}$ direction, while the $\mathbf{\hat{t}} \mathbf{\hat{t}}$ term provides a similar correction to the drag in $\mathbf{\hat{t}}$ to what it did in the SBT case.

\section{A numerical implementation of slender-ribbon theory} \label{sec:numerics}

Our final integral formulation, Eq.~\eqref{SREs}, allows for  investigations of the dynamics of slender ribbons at low Reynolds number. If the force distribution is known along the  ribbon plane, it is very easy to calculate the resultant motion by integration. However, in most cases of practical interest, the problem requires an inversion: it is the motion which  is known and the force distribution $\mathbf{f}$ needs to be computed by inverting the integrals. This is typically done numerically.

Different computational methods may be used for the inversion, and here we employ a Galerkin method \cite{Kloeden1992a}. The force distribution is expanded in terms of an infinite set of orthogonal functions. The orthogonality of these functions is then  used to reduce the integral equation into an infinite set of linear equations. Truncating and solving the remaining equations gives an approximation to the force distribution. The Galerkin method allows to employ similar simplifications to those used in  SBT  \cite{Gotz2000}.

The first integral in Eq.~\eqref{SREs} must be divided into two parts to implement these simplifications: one integral which has known eigenfunctions (Legendre polynomials) and the remaining behavior. With this in mind, we rewrite the integral result  as
\begin{eqnarray}\label{49}
\notag 8 \pi \mathbf{U}(s_1,s_2) &=&    \int_{-1}^{1} \,d t_{1}  \left[ \frac{(\mathbf{I} +\mathbf{\hat{R}_{0}} \mathbf{\hat{R}_{0}} )}{|R_{0}|}  -\frac{  (\mathbf{I} +\mathbf{\hat{t}} \mathbf{\hat{t}} )}{|t_{1}-s_{1}|}\right]\cdot \rho_{1}(t_{1}) \left\langle\mathbf{f}\right\rangle(t_{1}) \\
\notag&& +  (\mathbf{I} +\mathbf{\hat{t}} \mathbf{\hat{t}} ) \cdot \int_{-1}^{1} \,d t_{1}  \left[\frac{ \rho_{1}(t_{1})\left\langle\mathbf{f}\right\rangle(t_{1})  -\rho_{1}(s_{1})  \left\langle\mathbf{f}\right\rangle(s_{1})}{|t_{1}-s_{1}|}\right] \\
\notag &&+ \rho_{1} (s_{1}) \left[ L_{SRT} (\mathbf{I} +\mathbf{\hat{t}} \mathbf{\hat{t}}) 
-2 \mathbf{\hat{t}} \mathbf{\hat{t}} + 2 \mathbf{\hat{T}} \mathbf{\hat{T}}  \right] \cdot \left\langle\mathbf{f}\right\rangle(s_{1}) \\&&
+ \rho_{1}  (\mathbf{I} +\mathbf{\hat{t}} \mathbf{\hat{t}}) \cdot \int_{-1}^{1} \,d t_{2}  \log\left(\frac{1}{|t_{2}-s_{2}|^{2}}\right)   \mathbf{f}(s_{1},t_{2}),
\end{eqnarray}
where $L_{SRT} = \log\left[\frac{4(1-s_{1}^{2})}{b_{\ell}^{2} \rho_{1}^{2}}\right]$. The second integral in Eq.~\eqref{49} has eigenfunctions of Legendre polynomials, i.e.
\begin{equation}
 \int_{-1}^{1} \,d t_{1}  \left[\frac{ P_{n}(t_{1})  -P_{n}(s_{1}) }{|t_{1}-s_{1}|}\right] = - L_{n} P_{n}(s_{1}),
\end{equation}
where $P_{n}(s_{1})$ is the Legendre polynomial of order $n$, $L_{0} =0$  and $L_{n} = \sum_{i=1}^{n} 1/i$ for $n>0$ \cite{Gotz2000}. This suggests that $ \rho_{1} \left\langle\mathbf{f}\right\rangle$ should be expanded as a series of Legendre polynomials. We define $\mathbf{g}(s_{1},s_{2}) = \rho_{1}(s_{1}) \mathbf{f}(s_{1},s_{2})$ to absorb the $\rho_{1}$ dependence into $\mathbf{f}$. Similarly to the choice in $s_{1}$, the orthogonal functions for the $s_{2}$ expansion should be chosen to simplify the integrals. Here again Legendre polynomials are chosen. The Legendre polynomials in $s_{2}$ simplify the calculation of $\left\langle\mathbf{f}\right\rangle$, the rigid body motions, the total force and the total torque on the body.  We thus write $\mathbf{g}(s_{1},s_{2})$ as an infinite sum of Legendre polynomial in $s_{1}$ and $s_{2}$ as
\begin{eqnarray}
\mathbf{g}(s_{1},s_{2}) &=& \sum_{i=0}^{\infty} \sum_{j=0}^{\infty} \mathbf{g}_{i,j} P_{i}(s_{1}) P_{j}(s_{2}),  \nonumber \\ 
&=& \left( P_{0}(s_{1}) \mbox{ } P_{1}(s_{1}) \mbox{ } P_{2}(s_{1}) \mbox{ } \cdots  \right)\cdot\left(
\begin{array}{ c c c c }
\mathbf{g}_{0,0} & \mathbf{g}_{0,1} & \mathbf{g}_{0,2} & \cdots \\
\mathbf{g}_{1,0} & \mathbf{g}_{1,1} & \mathbf{g}_{1,2} & \cdots \\
\mathbf{g}_{2,0} & \mathbf{g}_{2,1} & \mathbf{g}_{2,2} & \cdots \\
\vdots & \vdots & \vdots& \ddots
\end{array} \right)\cdot\left(
\begin{array}{ c}
P_{0}(s_{2}) \\
P_{1}(s_{2}) \\
P_{2}(s_{2}) \\
\vdots
\end{array} \right),\nonumber \\
& =& \mathbf{S_{1}}^{T}(s_{1}) \cdot \mathbf{G} \cdot \mathbf{S_{2}}(s_{2}),
\end{eqnarray}
where the $\mathbf{g}_{i,j}$ are constant three component vectors, the Latin indices ($i$, $j$) represent the Legendre polynomial order in ($s_{1}$,$s_{2}$) respectively, $\mathbf{G}$ is a matrix of the $\mathbf{g}_{i,j}$, $\mathbf{S_{1}}$ is a column vector of the $s_{1}$ orthogonal functions and $\mathbf{S_{2}}$ is a column vector of the $s_{2}$ orthogonal functions. 

Since Legendre polynomials satisfy the orthogonality condition 
\begin{equation}
\int_{-1}^{1} \,d t_{1}  P_{n}(t_{1}) P_{m}(t_{1}) = \frac{2}{2 n +1} \delta_{n,m},
\end{equation}
it is straightforward to show that
\begin{equation}
 \rho(s_{1}) \left\langle f_{\nu} \right\rangle = \int_{-1}^{1} P_{j}(s_{2}) \,ds_{2} G_{\nu; j,k} P_{k} (s_{1})=  2 \delta_{j,0} G_{\nu; j,k} P_{k} (s_{1}),
\end{equation}
where the Greek indices, $\nu$, corresponds to one of the Cartesian components, and repeated indices are summed over. The slender-ribbon equation can then be rewritten as 
\begin{eqnarray}
\notag 8 \pi U_{\eta}(s_{1},s_{2}) &= & 2 \delta_{j,0} G_{\nu; j,k} \int_{-1}^{1} \,d t_{1}  \left[ \frac{(\delta_{\eta,\nu} +\hat{R}_{0;\eta} \hat{R}_{0;\nu} )}{|R_{0}|}  -\frac{  (\delta_{\eta,\nu} +\hat{t}_{\eta} \hat{t}_{\nu} )}{|t_{1}-s_{1}|}\right] P_{k}(t_{1}) \\&&
\notag+ 2 (\delta_{\eta,\nu} +\hat{t}_{\eta} \hat{t}_{\nu} ) \delta_{j,0} G_{\nu; j,k} \int_{-1}^{1} \,d t_{1}  \left[\frac{P_{k}(t_{1}) -  P_{k}(s_{1})}{|t_{1}-s_{1}|}\right] \\ &&
\notag+ (\delta_{\eta,\nu}+\hat{t}_{\eta} \hat{t}_{\nu} )  G_{\nu; j,k} P_{k} (s_{1})  \int_{-1}^{1} \,d t_{2}  \log\left(\frac{1}{|t_{2}-s_{2}|^{2}}\right)  P_{j}(t_{2})\\
&& +  2\left[ L_{SRT} (\delta_{\eta,\nu} +\hat{t}_{\eta} \hat{t}_{\nu} ) -2 \hat{t}_{\eta} \hat{t}_{\nu} + 2 \hat{T}_{\eta} \hat{T}_{\nu}  \right] \delta_{j,0} G_{\nu; j,k} P_{k} (s_{1}).
\end{eqnarray}
Multiplying the above equation by $ P_{m} (s_{1})$ and $ P_{n}(s_{2})$ and integrating over all of $s_{1}$ and $s_{2}$ the equation reduces to
\begin{equation} \label{SRTnumeric}
8 \pi \xi_{\eta;n,m} 
=  G_{\nu; j,k}\left[ 4 \delta_{j,0} \delta_{n,0} \left(\beta^{a}_{\eta,\nu;k,m} + \beta^{b}_{\eta,\nu;k,m} +\beta^{d}_{\eta,\nu;k,m}  \right)  + \beta^{c}_{\eta,\nu;k,m} \Upsilon_{n,j}\right],
\end{equation}
where
\begin{eqnarray}
\xi_{\eta ;n,m} &=&  \int_{-1}^{1} \,ds_{1} \int_{-1}^{1} \,ds_{2}   P_{m} (s_{1})   P_{n}(s_{2}) U_{\eta}(s_{1},s_{2}), \\
\beta^{a}_{\eta,\nu ;k,m} &=& \int_{-1}^{1} \,d s_{1 }   P_{m}(s_{1}) \int_{-1}^{1} \,d t_{1}  \left(\frac{(\delta_{\eta,\nu} +\hat{R}_{0;\eta} \hat{R}_{0;\nu} )}{|R_{0}|}  -\frac{  (\delta_{\eta,\nu} +\hat{t}_{\eta} \hat{t}_{\nu} )}{|t_{1}-s_{1}|}\right) P_{k}(t_{1}), \\
\beta^{b}_{\eta,\nu;k,m} &=& \int_{-1}^{1} \,d s_{1 }  (\delta_{\eta,\nu} +\hat{t}_{\eta} \hat{t}_{\nu} )  P_{m} (s_{1})  \int_{-1}^{1} \,d t_{1}  \frac{P_{k}(t_{1}) -  P_{k}(s_{1})}{|t_{1}-s_{1}|}, \nonumber   \\
&=& - L_{k}   \beta^{c}_{\eta,\nu ;k,m}, \\
  \beta^{c}_{\eta,\nu ;k,m}  &=&\int_{-1}^{1} \,d s_{1 }    P_{m} (s_{1})  (\delta_{\eta,\nu} +\hat{t}_{\eta} \hat{t}_{\nu} )   P_{k} (s_{1}),   \\
\beta^{d}_{ \eta,\nu ;k,m} &=& \int_{-1}^{1} \,d s_{1 }   P_{m} (s_{1}) \left( L_{SRT} (\delta_{\eta,\nu} +\hat{t}_{\eta} \hat{t}_{\nu} ) -2 \hat{t}_{\eta} \hat{t}_{\nu} + 2 \hat{T}_{\eta} \hat{T}_{\nu}  \right)  P_{k} (s_{1}),   \\
  \Upsilon_{n,j} &=&\int_{-1}^{1} \,ds_{2} P_{n}(s_{2}) \int_{-1}^{1} \,d t_{2}  \log\left(\frac{1}{|t_{2}-s_{2}|^{2}}\right)  P_{j}(t_{2}).
\end{eqnarray}
 The integrals listed above involve known functions and therefore are easily computed using MATLAB \cite{MATLAB2014}. Care must be taken with $\beta^{a}_{\eta,\nu ;k,m}$ and $ \Upsilon_{n,j}$ as the individual terms in the integrand blow up at $t=s$, though the full integrals are non-singular. MATLAB handles this using quadrature methods, and so approximates the integral without experiencing sampling issues. Interestingly the $\Upsilon_{n,j}$ is the only integral relating to $s_{2}$ and it has no explicit dependence on the  shape of the ribbon. It is therefore possible to evaluate $\Upsilon_{n,j}$ once and use it for many different shape configurations.

 For the $\xi_{\eta;n,m}$ integral, further simplifications are possible in the case of rigid-body motions. In that situation, the body translates at constant velocity, $\mathbf{u}$, and rotates with  constant angular velocity, $\boldsymbol{\omega}$.   Since the equations are linear,  translation and rotation can be treated separately and combined at the end. Separating $\xi_{l;n,m}$ into rigid translation, $\xi_{\eta;n,m}^{\mathbf{u}}$, and rigid rotation, $\xi_{\eta;n,m}^{\boldsymbol{\omega}} $, the integral becomes
 \begin{eqnarray}
 \xi_{\eta;n,m}^{\mathbf{u}} &=& \int_{-1}^{1} \,ds_{1} \int_{-1}^{1} \,ds_{2}   P_{m} (s_{1})   P_{n}(s_{2}) u_{\eta}, \\
  \xi_{\eta;n,m}^{\boldsymbol{\omega}} &=& \int_{-1}^{1} \,ds_{1} \int_{-1}^{1} \,ds_{2}  \epsilon_{\eta \nu \upsilon} P_{m} (s_{1})   P_{n}(s_{2}) X_{\nu}(s_{1},s_{2}) \omega_{\upsilon},
 \end{eqnarray}
 where $\epsilon_{\eta \nu \upsilon}$ is the levi-civita symbol and $X_{\nu}(s_{1},s_{2})$ is the $\nu$th component of the ribbon plane position vector. Using Eq.~\eqref{scaledsheet} and evaluating these integrals as far as possible, they become
  \begin{eqnarray}
 \xi_{\eta;n,m}^{\mathbf{u}} &=& 4 \delta_{m,0} \delta_{n,0} u_{\eta}, \\
  \xi_{\eta;n,m}^{\boldsymbol{\omega}} &=&\epsilon_{\eta \nu \upsilon}  \omega_{\upsilon}\left[ 2 \delta_{n,0} \int_{-1}^{1} \,ds_{1}    P_{m} (s_{1}) r_{\nu}(s_{1}) + \frac{2}{3} b_{\ell} \delta_{n,1} \int_{-1}^{1} \,ds_{1}   P_{m} (s_{1})  \rho_{1}(s_{1})  \hat{T}_{\nu}(s_{1}) \right].
 \end{eqnarray}
 This significantly simplifies the integrals which need to be computed to obtain $\xi$. Similarly, the total force and torque can be simplified to
\begin{eqnarray}
  \mathbf{F} &=&  4 \mathbf{g}_{0,0}, \\
  \mathbf{L} &=& \sum_{i,j} \int_{-1}^{1} \,dt_{1}\int_{-1}^{1} \,dt_{2}  \left( \mathbf{r}(t_{1}) + b_{\ell} t_{2} \rho_{1}(t_{1})  \mathbf{\hat{T}}(t_{1}) \right)\times  \mathbf{g}_{i,j} P_{i}(t_{1}) P_{j}(t_{2}), \nonumber \\
  &=& \sum_{i} \left(2 \int_{-1}^{1} \,dt_{1} P_{i}(t_{1}) \mathbf{r}(t_{1}) \times  \mathbf{g}_{i,0}  + \frac{2}{3} b_{\ell} \int_{-1}^{1} \,dt_{1}   P_{i}(t_{1}) \rho_{1}(t_{1})  \mathbf{\hat{T}}(t_{1}) \times  \mathbf{g}_{i,1}   \right).
  \end{eqnarray}

The Galerkin system, Eq.~\eqref{SRTnumeric}, leads to an infinite set of linear equations. For numerical work these equations need to be truncated. 
The number of orthogonal functions kept in $s_{1}$ and $s_{2}$ will be denoted  $N_{1}$ and $N_{2}$, respectively. The $\mathbf{g}$ coefficients can then be solved for by representing the tensors in Eq.~\eqref{SRTnumeric} by vectors and matrices and using standard matrix inversion. To do this the vector structure is divided into two levels: outer and inner. The full vector is divided into $N_{2}$ outer levels, and each outer level is further divided into $N_{1}$ inner levels. These levels are formatted such that the $i$th inner level in the $j$th outer level contains $\mathbf{g}_{i,j}$. Therefore subsequent inner levels, in a certain outer level, represent the different Legendre polynomials in $s_{1}$ (changing $i$), while the subsequent outer levels represent the different Legendre polynomials in $s_{2}$ (changing $j$). These vectors are of length $3 N_{1} N_{2}$. Thus the corresponding matrices are of size $9 N_{1}^{2} N_{2}^{2}$. This can be very large, however the terms needed to construct these matrices are defined by the integrals above and so can be evaluated separately and stored in small matrices rather than one large matrix.

\section{Validation of slender-ribbon theory:  plate ellipsoids} \label{sec:plateellipsoid}

The simplest structure with known mobility coefficients which can be modeled using SRT is a thin flat ellipsoid (plate ellipsoid). The resistance matrix for an arbitrary ellipsoid is known exactly \cite{LAMB1932,Perrin1934} and when said ellipsoid is sufficiently flat and long, it approaches the  slender-ribbon limit. We use these exacts results to validate our SRT approach.

The  force $F$ applied by an arbitrary ellipsoid, with semi-axes lengths \{$k$, $m$, $n$\}, on the fluid when translating at speed $U$ in the $k$ direction is given by
\begin{equation} \label{ellipseforce}
\frac{F}{ \pi \mu U} = \frac{16 }{\phi + \zeta_{k} k^{2}},
\end{equation}
while the torque applied on a fluid due to rotation with rate $\Omega$ around the $k$ direction is
\begin{equation}\label{ellipsetorque}
\frac{L}{\pi \mu \Omega} = \frac{16}{3} \frac{m^{2} +n^{2}}{m^{2} \zeta_{m} + n^{2} \zeta_{n}},
\end{equation}
where we have 
\begin{eqnarray}
\phi &=& \int_{0}^{\infty}\,d x \frac{1}{\sqrt{(k^{2}+x)(m^{2}+x)(n^{2}+x)}}, \\
\zeta_{k} &=& \int_{0}^{\infty}\,d x \frac{ 1}{(k^{2}+x)\sqrt{(k^{2}+x)(m^{2}+x)(n^{2}+x)}}, \\
\zeta_{m} &=& \int_{0}^{\infty}\,d x \frac{ 1}{(m^{2}+x)\sqrt{(k^{2}+x)(m^{2}+x)(n^{2}+x)}}, \\
\zeta_{n} &=& \int_{0}^{\infty}\,d x \frac{ 1}{(n^{2}+x)\sqrt{(k^{2}+x)(m^{2}+x)(n^{2}+x)}}\cdot
\end{eqnarray}
 These integrals can  be easily evaluated using MATLAB.

\begin{figure}[p]
\begin{center}
\includegraphics[width=.9\textwidth, , bb= 0 0 460 558]{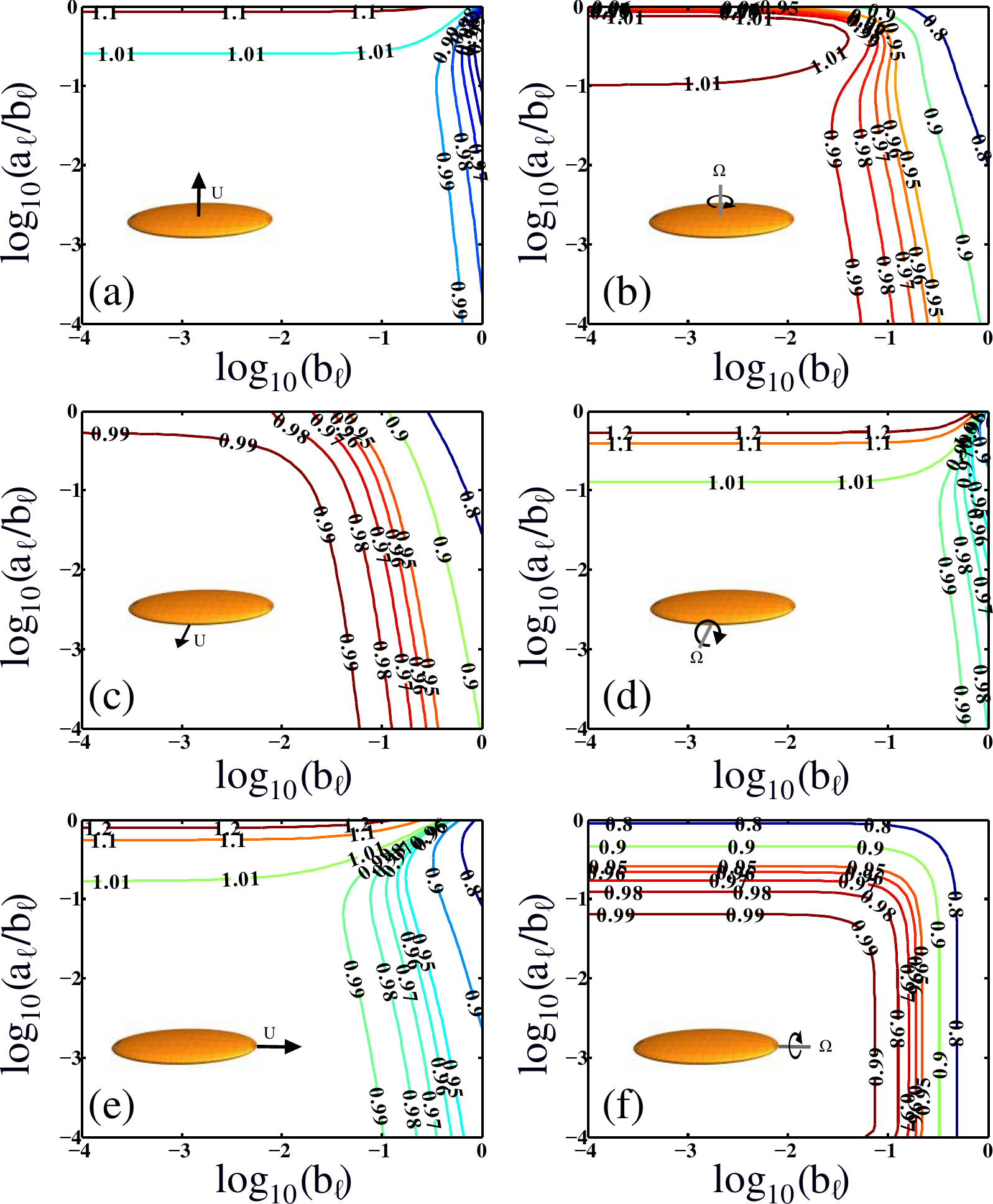}
\caption{Rigid-body motion of a plate ellipsoid:  ratio between the net force or torque    computed using  SRT  and the exact solution. 
The figures shows  contour levels of the ratios for different values of $b_{\ell}$ and $a_{\ell}/b_{\ell}$ in six distinct cases: (a)   ratio for the force in the $a$ direction from translation in the same direction; (b)   ratio for the torque in the $a$ direction from rotation in the same direction; (c) ratio for the force in the $b$ direction from translation in the same direction; (d)  ratio for the torque in the $b$ direction from rotation in the same direction; (e)  ratio for the force in the $\ell$ direction from translation  in the same direction; (f)   ratio for the torque in the $\ell$ direction from rotation in the same direction. The SRT computations were carried out  using   $N_{1} =15$ and $N_{2}=15$.}
\label{fig:plateellipse}
\end{center}
\end{figure}

The parametrisation of a flat ellipsoidal ribbon has a straight centerline, $\mathbf{r}(s_{1}) = s_{1}\mathbf{\hat{x}}$, a constant plane vector, $\mathbf{\hat{T}} = \mathbf{\hat{y}}$, and $\rho_{1} = \sqrt{1-s_{1}^{2}}$. The ribbon plane obtained from this parametrisation is shown below in 
Fig.~\ref{fig:twistellipse}a. For the plate ellipsoid many of the integrals simplify analytically and only the $\Upsilon_{n,j}$ and $\xi_{l;n,m}$ integrals need to be computed numerically. Two parameters may be varied to compare SRT to the exact solution: $b_{\ell}$ and $a_{\ell}/b_{\ell}$. 

We plot in Fig.~\ref{fig:plateellipse} iso-contours for the ratio between the hydrodynamic resistances obtained using SRT to the exact resistances of a plate ellipsoid, for both translation and rotation in all three directions. 
The figure shows that SRT converges to within 1\% of the exact solution rapidly as both $b_{\ell}$ and $a_{\ell}/b_{\ell}$ decrease. Recall that the asymptotic limit in which SRT is expected to be valid is $b_{\ell}\ll 1$ and 
$a_{\ell}/b_{\ell} \ll 1$. We note that the value of $b_{\ell}$ tends to have a larger affect on the accuracy than that of $a_{\ell}/b_{\ell}$. The convergence rate differs for the different force and torque components but all have converged to within 1\% error by $b_{\ell} = a_{\ell}/b_{\ell} = 10^{-2}$. The plots also reveal that the torque terms converge without the need for rotlet singularities (which would be needed, for example, for the rotation of a spherical body). In the ribbon case, the rotation of a sheet of stokeslets can thus adequately capture the leading-order torque.

\section{Comparison with computations for ribbon helices}\label{sec:keaveny}

The dynamics of thick ribbons twisted into helices has been explored numerically previously using a boundary integral  method \cite{Keaveny2011,Keaveny2013}.  
These ribbons were unfortunately not slender, and the thinnest ribbon studied had $b_{\ell} =1/25$ and $a_{\ell}/b_{\ell} =1/4$. From Sec.~\ref{sec:plateellipsoid} and Fig.~\ref{fig:plateellipse}, we  see that the error in the resistance coefficients obtained using SRT for a plate ellipsoid with the same dimensions is up to 10\%, and thus we should expect results with errors of a similar order of magnitude when comparing SRT with the work in Refs.~\cite{Keaveny2011,Keaveny2013}. 

\begin{figure}
\begin{center}
\includegraphics[width=.8\textwidth, bb= 0 0 569 240]{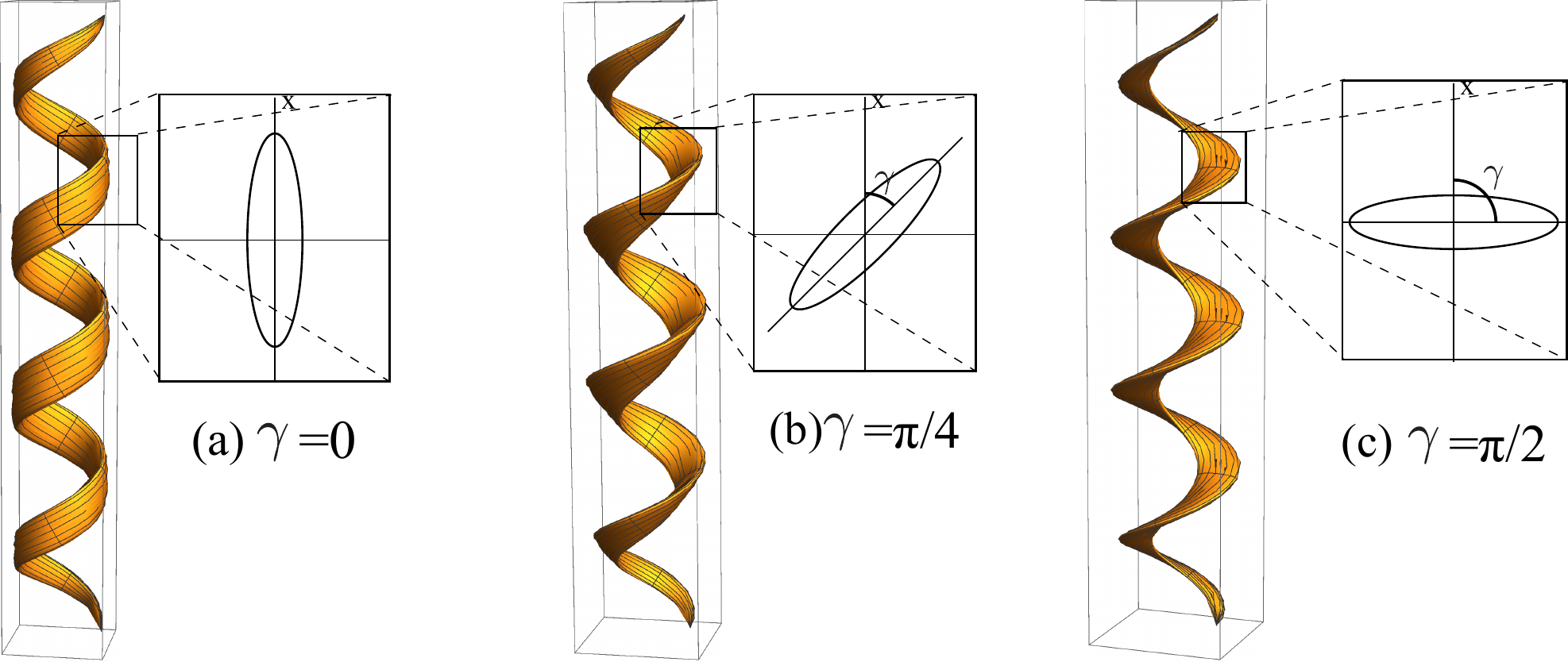}
\caption{A sample of the helical ribbon-like structures explored computationally in Ref.~\cite{Keaveny2011}. The parameter $\gamma$ measures the relative angle between the ribbon plane and the central axis of the helix. These helices have a cross sectional aspect ratio of 4.}
\label{fig:helixrib}
\end{center}
\end{figure}

In Fig.~\ref{fig:helixrib} we show a sample of the shapes explored in Ref.~\cite{Keaveny2011}. The parametrisation used in that paper was
\begin{eqnarray}
\mathbf{r}_{h}(s_{1}) &=& \left\{\beta_{h} \cos(k s_{1}),\beta_{h} \sin(k s_{1}), \alpha_{h} s_{1}  \right\}, \label{helixcent} \\
\mathbf{\hat{T}}_{h}(s_{1}) &=& \cos(\gamma) \mathbf{\hat{b}}_{h} -\sin(\gamma) \mathbf{\hat{n}}_{h}, \label{helixT}
\end{eqnarray}
where $\mathbf{r}_{h}(s_{1})$ is the centerline, $\beta_{h}$ is the helix radius, $k$ is the wavenumber, $\alpha_{h}$ is the cosine of the helix angle, $\gamma$ is the angle between the central axis and the ribbon plane (illustrated in Fig.~\ref{fig:helixrib}), and $\mathbf{\hat{n}}_{h}$ and $\mathbf{\hat{b}}_{h}$ are the normal and bi-normal vectors to the helix centerline. This helix parametrisation relates $\alpha_{h}$, $\beta_{h}$ and $k$, through $\alpha_{h}^{2} +\beta_{h}^{2} k^{2} =1$, and the length measured along the helix axis is related to the centerline length by $2 L_{axis} = 2 \alpha_{h} \ell$.  In the work from Ref.~\cite{Keaveny2011}, $\rho_{1}$ was taken to be $\sqrt{1-s_{1}^{2}}$ and the cross-sectional shape was an ellipse with $\mathbf{\hat{T}_{h}}$ pointing to the major axis.   The simulations in Ref.~\cite{Keaveny2011} considered ribbons with cross-sectional aspect ratios of 1, 2 and 4 while keeping the cross sectional area constant. The computational results showed that the ribbon propelling the quickest for a set external torque (velocity per unit torque) is similar to an Archimedean screw ($\gamma = \pi/2$, Fig.~\ref{fig:helixrib}c) while the slowest  is the structure shown in Fig.~\ref{fig:helixrib}a ($\gamma = 0$). 

\begin{figure}
\begin{center}
\includegraphics[width=.5\textwidth, bb= 0 0 264 206]{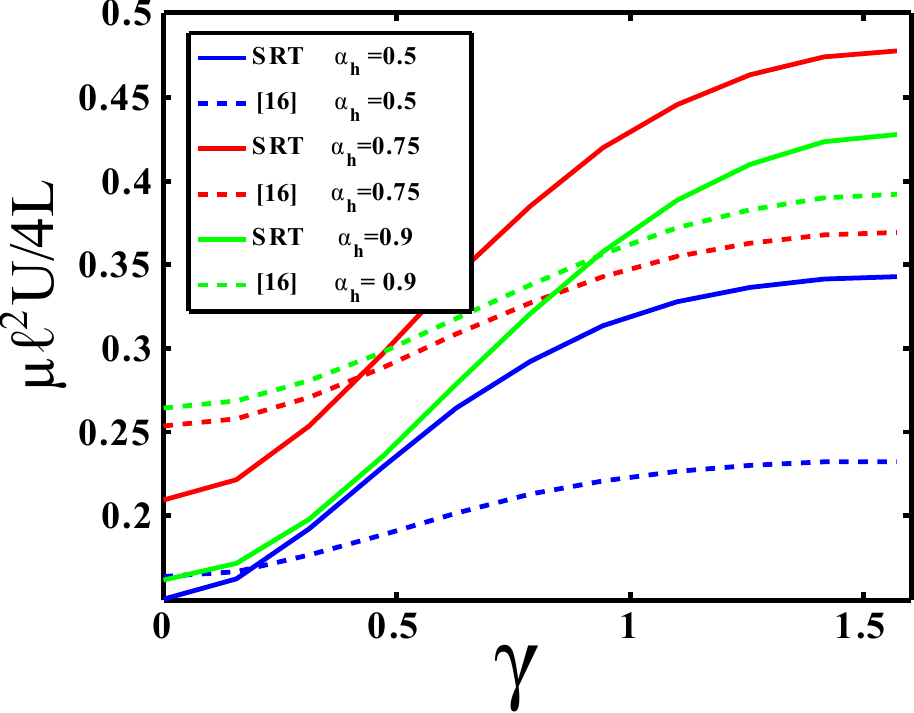}
\caption{Scaled force-free velocity per unit torque for a ribbon helix with different values of $\gamma$. The SRT results are displayed in sold lines while the  numerical results of Ref.~\cite{Keaveny2011} are shown in dashed line. The SRT numerics are done with $N_{1} =35$ and $N_{2}=15$.}
\label{fig:helixribcomp}
\end{center}
\end{figure}

We compare the computational results of Ref.~\cite{Keaveny2011} with aspect ratio $4$ with those of SRT.  The  parametrisation above is used with $k=4 \pi$, $b_{\ell} =1/25$ and $\alpha_{h} =0.5, 0.75$ and $0.9$. A net torque, $L$, is applied along the axis of the helix and we compute the resulting translational velocity, $U$. The comparison is shown in  Fig.~\ref{fig:helixribcomp} with SRT results in solid lines while the computational results from Ref.~\cite{Keaveny2011} are plotted in dashed lines. Qualitatively, the results display the  same dependence on $\gamma$. When $\gamma =\pi/2$ the velocity per unit torque is at a maximum while when $\gamma=0$ it is at a minimum. Quantitatively,  SRT overestimates the results near $\gamma=\pi/2$ and underestimates them near $\gamma=0$.  The increase in propulsion at $\gamma =\pi/2$ is likely to be due to  the body being thinner in the SRT framework, and thus experiencing  less drag. The cause of the decrease at $\gamma=0$ is unknown but is consistent with the behaviour of $\gamma=0$ found in Ref.~\cite{Keaveny2011} as the aspect ratio increased.

\section{Comparison with experiments for ribbon microswimmers} \label{sec:zhang}

\begin{figure}
\begin{center}
\includegraphics[width=.6\textwidth, bb= 0 0 567 254]{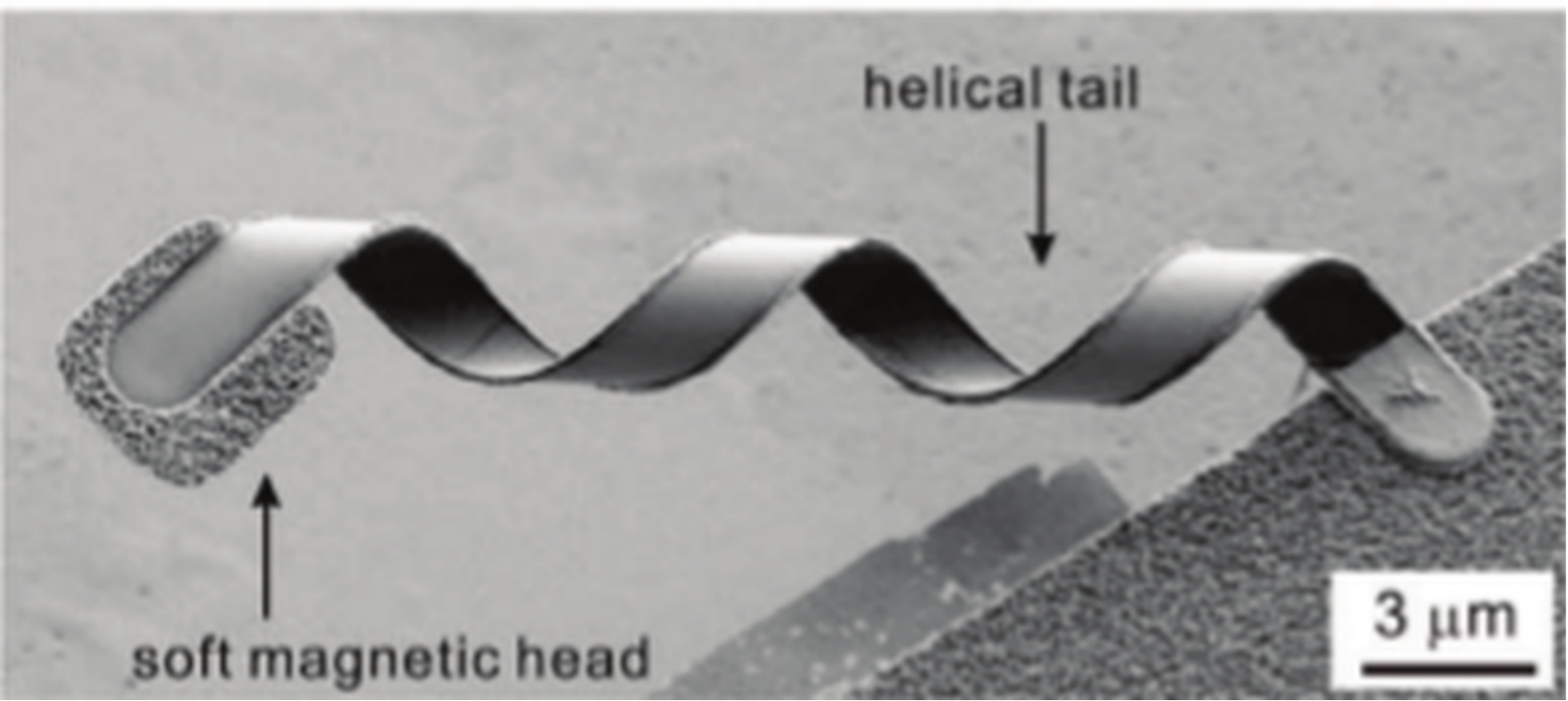}
\caption{A scanning electron microscope micrograph of an  artificial bacterial flagellum with a diameter of 2.8~$\mu$m~\cite{Zhang2009}. Adapted  from L. Zhang, J. J. Abbott, L. Dong, K. E. Peyer, B. E. Kratochvil, H. Zhang, C. Bergeles, and
B. J. Nelson, ``Characterizing the swimming properties of artifcial bacterial 
agella.," \textit{Nano Lett.},  \textbf{9}, 3663 (2009), Copyright 2009 American Chemical Society.}
\label{fig:fabricationphoto}
\end{center}
\end{figure}

Slender ribbons have been used to devise micron-scale artificial swimmers  termed {\it  artificial bacterial flagella} \cite{Zhang2009,Zhang2010} .  These swimmers consist of a magnetic head and a thin ribbon tail twisted into a helical shape similarly to what would be formed by a straight ribbon twisted around a pencil (Fig.~\ref{fig:fabricationphoto}). These microswimmers are then rotated through the use of an external rotating  magnetic field, which leads to forward propulsion. The ribbons used in these studies were typically tens of $\mu$m long, a few $\mu$m wide, and tens of nm thick, leading to $b_{\ell} = O(0.1)$ and $a_{\ell}/b_{\ell} =O(0.01)$, an appropriate dimensionless limit to address using slender-ribbon theory. This application is done with some caution as Fig.~\ref{fig:fabricationphoto} suggests that the radius curvature of the ribbon may be of a similar order to $b_{\ell}$.

Specifically, the swimmers in Ref.~\cite{Zhang2009} were made with a ribbon with dimensions $2b=1.8$~$\mu$m and $2a=42$~nm together with a square  magnetic head with dimensions of $4.5$~$\mu$m$\times 4.5$~$\mu$m $\times 200$~nm.   At the time of the experiment the swimmer has 4.5 wavelengths along its body and a length relative to the helix axis of $L_{axis}=38$~$\mu$m. Before twisting, the ribbon was $49.7$~$\mu$m long, and the swimmer had a helix diameter of $2.8$~$\mu$m immediately after fabrication. However, these helical dimensions are inconsistent with each other, which is likely to be due to the swimmers slowly changing  dimensions for a few weeks after fabrication (Zhang \& Nelson, private communication and Ref.~\cite{Zhang2009a}). Therefore, the helix diameter and centerline length at the time of the experiment are not exactly known. To address this issue, we  run two different simulations, one where the centerline length is taken to be $2\ell=49.7$~$\mu$m, and one where the helix diameter is taken to be $2\beta_{h}=2.8$~$\mu$m. We expect the true dimensions to be somewhere in between these two values. 

The head and the twisted ribbon are treated separately in our model of the microswimmer, with no hydrodynamic interactions. The resistance coefficients of the full swimmer is then the sum of resistance coefficients on the head and the ribbon. This leaves the coupling coefficient as that of the helical ribbon while changing the other coefficients. The head is modeled as an oblate spheroid aligned such that the shortest direction is perpendicular to the helix axis of the body.  The resistance coefficients on this spheroid can then be calculated using Eqs.~\eqref{ellipseforce} and \eqref{ellipsetorque}. Assuming a dynamic  viscosity of water of $10^{-3}$~Pa.s, the drag from translation on the head is found to be $2.49\times 10^{-8}$~N.s.m$^{-1}$ and the torque from rotation is $1.22\times 10^{-19}$~N.s.m. The ribbon is assumed to have the form given by Eqs.~\eqref{helixcent} and \eqref{helixT}, where $\gamma$ has been set to 0. This description is then used to compute the resistance coefficients for helices with $4.5$ waves along their length, and an axial length of $2L_{axis}=38$~$\mu$m. As  mentioned above, two separate  cases are considered: one where $2\ell=49.7$~$\mu$m and one where $2\beta_{h}=2.8$~$\mu$m. 

\begin{figure}[b]
\begin{center}
\includegraphics[width=.6\textwidth, bb= 0 0 545 379]{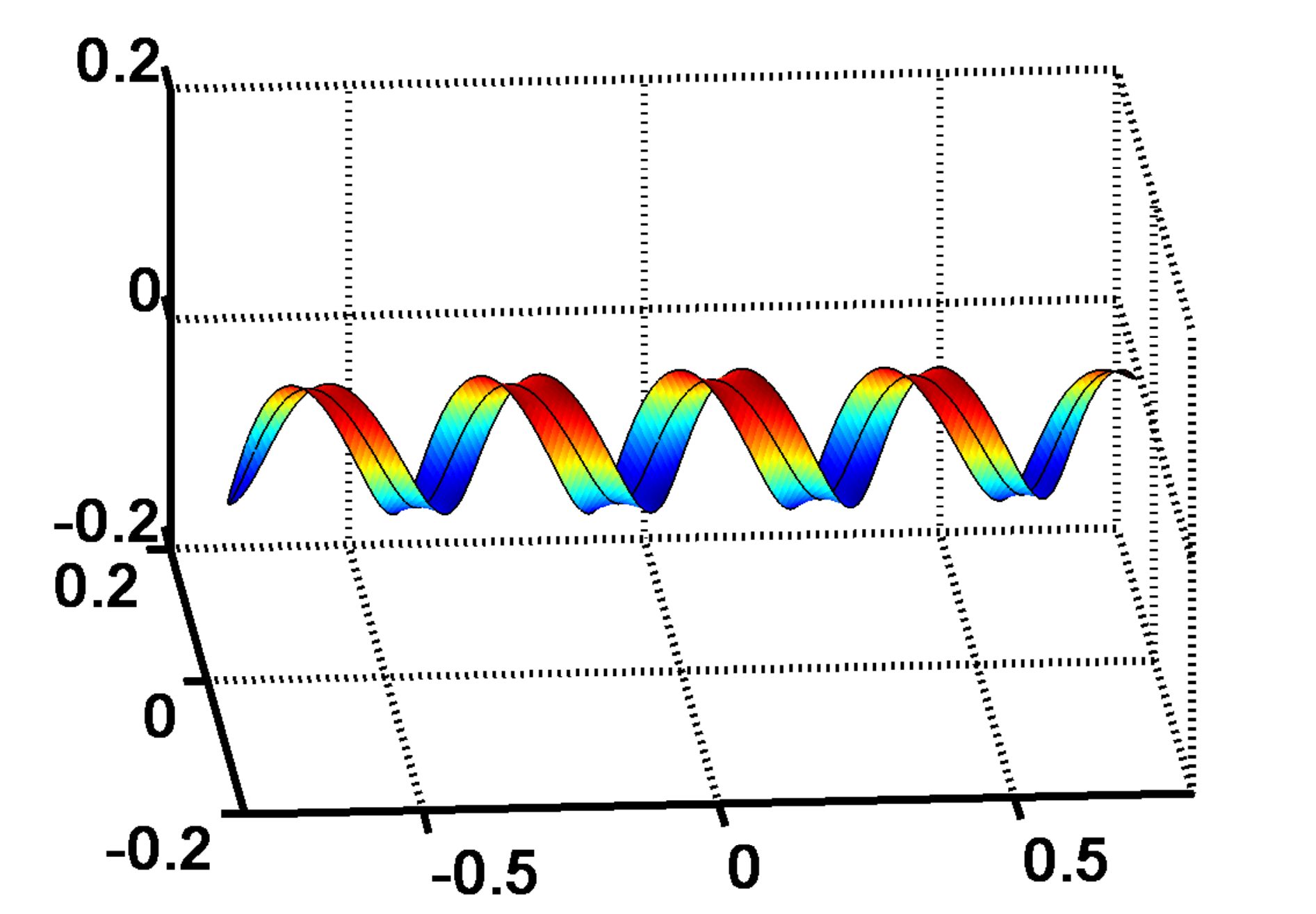}
\caption{Shape of the ribbon used to model the artificial micron-scale swimmer from Ref.~\cite{Zhang2009}. In this case $2\beta_{h}$ has been taken to be $2.8$~$\mu$m. All lengths are scaled by the half centerline length $\ell$.}
\label{fig:modelswimm}
\end{center}
\end{figure}

In Fig.~\ref{fig:modelswimm} we show the parametrization of the ribbon plane used in  the case $2\ell=49.7$~$\mu$m. 
 In  Ref.~\cite{Zhang2009}, the coefficients of the  resistance matrix  for such a helical swimmer were characterized experimentally, with values  listed in the first column of Table~\ref{tab:helixswimmer} while the results for both cases  $2\beta_{h}=2.8$~$\mu$m and $2\ell=49.7$~$\mu$m  are shown in the second and third column, respectively.  In this table, $A$ denotes the hydrodynamic resistance coefficient relating the drag force experienced  parallel to the helical axis from translation in the same direction, which  is therefore composed of the drag on the helix and the drag on the head; $B$ is the hydrodynamic force experienced  parallel to the helix axis from rotation around the helix axis; finally $C$ is the hydrodynamic torque experienced around the helical axis from rotation around said axis of both the head and the helix. We also give the value of the ratio $B/A$, which  is  important  in the context of the locomotion. Indeed, since the swimmer must be force free, the translational  velocity per unit angular velocity is given by the negative of the coupling divided by the translational drag ($-B/A$), the values of which are  displayed on the last  row of Table~\ref{tab:helixswimmer}.

\begin{table}[t]
\begin{center}
\begin{tabular}{| c| c| c |c | c|} \hline
 & \quad Experiments~\cite{Zhang2009}  \quad & \quad SRT ($2\beta_{h}=2.8$~$\mu$m) \quad  & \quad SRT ($2\ell=49.7$~$\mu$m) \quad & \quad Ref.~\cite{Keaveny2013}'s model \quad\\ \hline
\,\, $A$ ($10^{-7}$~N.s.m$^{-1}$) \,\,  & 1.5 & 1.04 & 0.973& 0.937\\
$B$  ($10^{-14}$~N.s) & -1.6  & -1.32 & -0.997& -1.63\\
$C$  ($10^{-19}$~N.m.s)& 2.3 & 6.81& 4.94 & 10.1 \\
$B/A$ ($10^{-7}$~m) & -1.07 & -1.27 & -1.02& -1.74 \\ \hline
\end{tabular}
\caption{Hydrodynamic resistance coefficients for ribbon microswimmer. 
Left:  Experimental measurements  \cite{Zhang2009}; Middle-Left: SRT results assuming   $2\beta_{h}=2.8$~$\mu$m; Middle-Right: SRT results with  $2\ell=49.7$~$\mu$m; Right: the model swimmer results from Ref.~\cite{Keaveny2013}. See text for the definitions of the resistance coefficients $A$, $B$ and $C$.  
The SRT computations are  done with $N_{1} =35$ and $N_{2}=15$.}
\label{tab:helixswimmer}
\end{center}
\end{table}

Inspecting the results in Table~\ref{tab:helixswimmer}, we see that both  cases give results close to the experimental measurements of Ref.~\cite{Zhang2009}. The geometry with  $2\beta_{h}=2.8$~$\mu$m gives results closer to the experiments  for force due to translation ($A$) and due to rotation ($B$) while the ratio $B/A$ and the torque resistance to rotation ($C$) is best approximated by the geometry with  $2\ell=49.7$~$\mu$m.   Therefore slender-ribbon theory, despite the fact that it is only valid asymptotically in the mathematically-slender limit, can be used as an accurate predictive tool to design microscopic swimmers similar to those from Refs.~\cite{Zhang2010,Zhang2009}.

 Table~\ref{tab:helixswimmer}  shows that the drag force, $A$, and the coupling force from rotation, $B$, of the model swimmer are smaller than the experimental values. This is likely due to the ellipsoidal cross section used to model the ribbon and head and the asymptotic treatment of the ribbon. The ribbon and head of the swimmer in Ref.~\cite{Zhang2009} are more square and have finite thickness (Fig.~\ref{fig:fabricationphoto}). Therefore they will have larger values of $A$ and $B$. 
 
 Furthermore, for both geometrical models,  the torque from rotation around the helix axis, $C$, is larger than that measured in Ref.~\cite{Zhang2009}. 
 This increase probably arises from the chose parametrization for the ribbon. As can be seen in Fig.~\ref{fig:modelswimm}, the edges of the ribbon  curve slightly. This curving is due to the twisting of the helix centerline. As the centerline curves around, the lines along $s_{2}$ fall such that when cut down the helix axis (not down $\mathbf{\hat{T}}$) the cross section appears curved.   
We anticipate that this curving of the edges  only significantly changes the value of $C$. Indeed, as  the slender-ribbon equations assume that the system consists of locally flat segments, this curving has essentially no influence on the value of  $A$. Similarly because $B$ quantifies  the torque induced by a translation, it follows roughly a linear surface dependence thereby giving $B$ a linear dependence on $s_{2}$.  The integration of $s_{2}$ over even bounds therefore removes any affect of said curving in $B$. In contrast, the torque from rotation, $C$, has quadratic dependence on the surface, and integration of an even function over even bounds does not cancel out. Hence this curving could significantly increase the value of $C$. Unfortunately because the current form of SRT requires $\mathbf{\hat{T}}\cdot \mathbf{\hat{t}}=0$ for all $s_{1}$, a parametrisation where the curving does not occur is not possible. But with a different mathematical approach to slender ribbons, this could perhaps be tackled. 

Previously these {\it  artificial bacterial flagella} have been modelled using a boundary integral formulation \cite{Keaveny2013}, with results given in the fourth column of table~\ref{tab:helixswimmer}. This model included interactions between the head and the tail and set $\alpha_{h}=0.7$. However, like the boundary integral work in Ref.~\cite{Keaveny2011}, $b_{\ell}/a_{\ell}=4$ and there was only four waves along the body centerline (including the head). Though the results of Sec.~\ref{sec:keaveny} show strong dependence on the configuration and $b_{\ell}/a_{\ell}$, Ref.~\cite{Keaveny2013}'s model closely replicated the coupling coefficent for the swimmer, $B$. This value is much closer than either SRT model. However the SRT models have closer values for the linear drag, $A$, the torque from rotation, $C$, and the velocity per unit angular velocity $-B/A$.

\section{Comparison  with slender body theory} 
\label{sec:sbtequiv}

There are, of course, many shapes and configurations that slender-ribbon theory can model which cannot be tackled using slender-body theory. For example, SBT cannot  replicate the behavior of a plate ellipsoid (Sec.~\ref{sec:plateellipsoid}) or the behavior created by a twisted ribbon (Sec.~\ref{sec:twistellipse}). However,  for example for ribbons with a helical centerline, one may ask how accurately the system could be represented by a slender body with the same centerline but a different effective thickness? Or how does the swimming speed of a slender body compare to that of a slender ribbon?
\begin{table}[ht]
\begin{center}
\begin{tabular}{|c| c| c |c|} \hline
 \multirow{2}{*}{\,\,Coefficients\,\,} & Microswimming   & SRT simulations of & \,\,SRT simulations for    \,\, \\
& experiments \cite{Zhang2010,Zhang2009} &  experiments \cite{Zhang2010,Zhang2009} &\,\, twisted ribbon\,\, \\ \hline
All & 8.80 $\times$10$^{-4}$ & 0.0192  & 0.0284\\ \hline
$A$ \& $B$ & 0.0418 & 0.0395 & 0.0241 \\ \hline
$A$ & 0.0800 &  0.0147 & 0.0262\\ \hline
$B$ &\,\, 0.0389 (4.37$\times$10$^{-4}$) \,\, & \,\,0.0411 (4.23$\times$10$^{-4}$)\,\,& 0.0126\\ \hline
$C$ & 4.80 $\times$10$^{-4}$& 0.0190 & 0.0287 \\ \hline
\end{tabular}
\caption{Radii of a slender body (normalized by half its centerline length) which best replicates the slender-ribbon coefficients with a helical centerline for three cases: the microswimming experiments \cite{Zhang2010,Zhang2009}, the SRT simulations for the same experiments, and the SRT simulations of a twisted ribbon with 
$\alpha_{h}=0.75$, $\gamma={\pi}/{2}$.  The resistance coefficients considered are: $A$,  drag along the helix axis from translation  along the same direction; $B$, torque around the helix axis from translation along the helix axis;  $C$,  torque around the helix axis from rotation around the helix axis. The ``All''  case considers  the configuration which minimizes the sum relative difference squared for the three terms while  ``$A$ \& $B$'' minimizes the sum of the relative difference squared for both $A$ and $B$. When there is more than one radii which replicates a given coefficient the largest is listed followed by the others in brackets. Note that the centerline and ribbon width are different between the experimental models and the $\alpha_{h}=0.75$ case.
}
\label{tab:sbtexp}
\end{center}
\end{table}

\begin{figure}
\begin{center}
\includegraphics[width=0.88\textwidth, bb = 0 0 572 424]{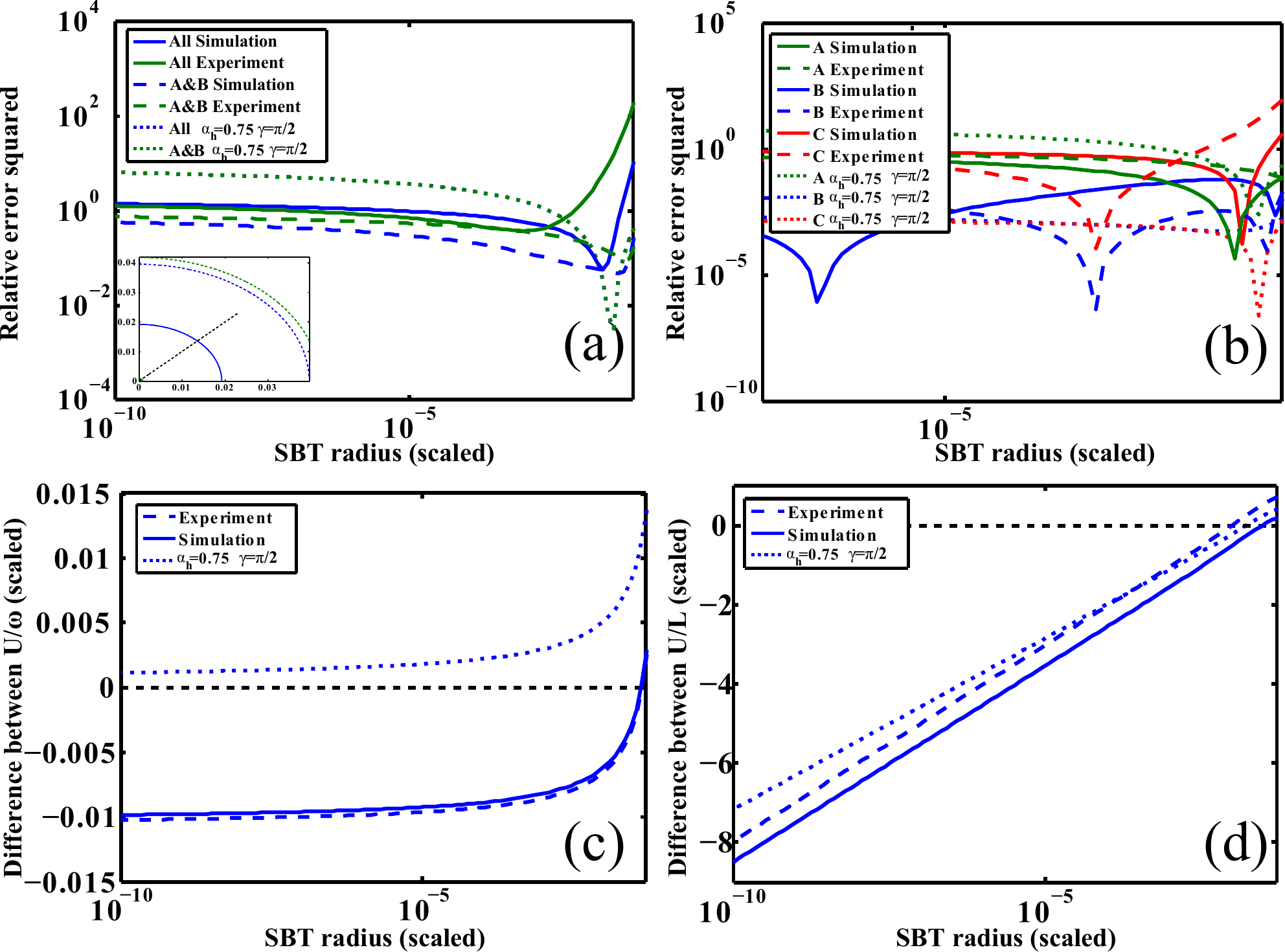}
\caption{Comparisons between a slender ribbon and a slender body with the same helical centerline as a function of the radius of the slender body scaled by half the centerline length.    Results are shown for the  microswimming experiments \cite{Zhang2010,Zhang2009}  (dashed), the SRT results of the same experiments  (solid), and the SRT results for a twisted ribbon with  $\alpha=0.75$ and $\gamma=\pi/2$ (dotted).  The plotted results are (a): the sum of the relative difference squared for all  coefficients $A$, $B$ and $C$ (blue) and  $A$ \& $B$ (green) with inset  depicting   the best radii in the microswimmer cases;  (b): the relative difference squared for $A$, $B$ and $C$ individually; (c): the difference between the velocity per unit rotation;  (d): the difference between the velocity per unit torque.  Numerical results obtained with $N_{1} =35$ and $N_{2}=15$.}
\label{fig:sbtexp}
\end{center}
\end{figure}

In this section we use a numerical implementation of SBT \cite{Koens2014} to compare ribbons twisted into helices to slender bodies with the same centerline. The comparison focuses on two features: (1) for what effective filament radius does SBT best replicates the resistance coefficients, $A$, $B$, and $C$, of the slender ribbon?; (2)  how does the velocity per unit rotation and the velocity per unit torque of a ribbon compare with slender bodies of different  radii? We take the slender body that best replicates each individual coefficient as the body which minimizes the relative difference squared between SRT and SBT results, 
defined as  $(1-\alpha_{SBT}/\alpha_{SRT})^2$  for any computed quantity of interest $\alpha$. 
 Similarly the one the best replicates the total (the ``All'' test) and both $A$ and $B$ (the ``$A$ \& $B$'' test) is the filament radius that minimizes the sum of the relative difference squared. For the velocity per unit rotation, or per unit torque, the difference between the velocities from SRT and SBT is used, and a negative value indicates faster propulsion for a  slender body than a slender  ribbon.

We carry out these  comparisons for the experimental and simulated results of the microscopic  swimmer discussed in Sec.~\ref{sec:zhang}   \cite{Zhang2009} and for  our simulations of helical ribbons with $\alpha_{h}=0.75$ and $\gamma=\pi/2$ discussed in Sec.~\ref{sec:keaveny}. The experimental configuration was assumed to have a centreline length of $49.7$~$\mu$m. Furthermore since the experimental results includes the head, the drag and torque of the equivalent oblate spheroid is removed from the resistance coefficients. 

In  Table~\ref{tab:sbtexp} we list the values of the best (as defined above) radii  for the slender-ribbon shapes while Fig.~\ref{fig:sbtexp} shows the relative difference squared, the difference in the velocity per unit rotation,  and the difference in the velocity per unit torque. The radii of the slender bodies are always scaled by half the centerline length. Sharp peaks in Fig.~\ref{fig:sbtexp}a and b correspond to points where the match between the slender body and slender ribbon is exact. 
The minima for the ``All'' and the ``$A$ \& $B$'' cases are not sharply peaked,  indicating that one slender body cannot exactly match the full  dynamics of the ribbon. In the ``All'' test, the average percentage error for each coefficient at the best slender body radius is  20\%, 7.9\% and 4\%, for the swimming experiments of Refs.~\cite{Zhang2010,Zhang2009},  
 the corresponding SRT results, and the SRT simulations of the twist ribbons with $\alpha_{h}=0.75$, $\gamma={\pi}/{2}$. Hence the slender-ribbon behavior is only poorly replicated by a single slender body.   Interestingly, the coupling coefficient, $B$, has  no peak when $\gamma=\pi/2$, indicating that the coupling for the Archimedean screw cannot be replicated by a slender body with the same centerline.   
In Figs.~\ref{fig:sbtexp}c and d we measure the difference between the velocity per unit rotation, $U/\omega$, and the velocity per unit torque, $U/L$, respectively. For the velocity per unit rotation there is a sharp decrease and then an approach to an asymptote as the body gets logarithmically thinner. In contrast,  for the velocity per unit torque,  the difference  decreases linearly as the body gets logarithmically thinner. For the twisted ribbon with  $\gamma=\pi/2$, the difference in $U/\omega$ is never  negative showing that, at a given angular velocity, no slender body can propel faster than a slender ribbon with an identical centerline.

\section{Hydrodynamics of a thin twisted plate ellipsoid} \label{sec:twistellipse}

\begin{figure}
\begin{center}
\includegraphics[width=0.8\textwidth, bb = 0 0 480 382]{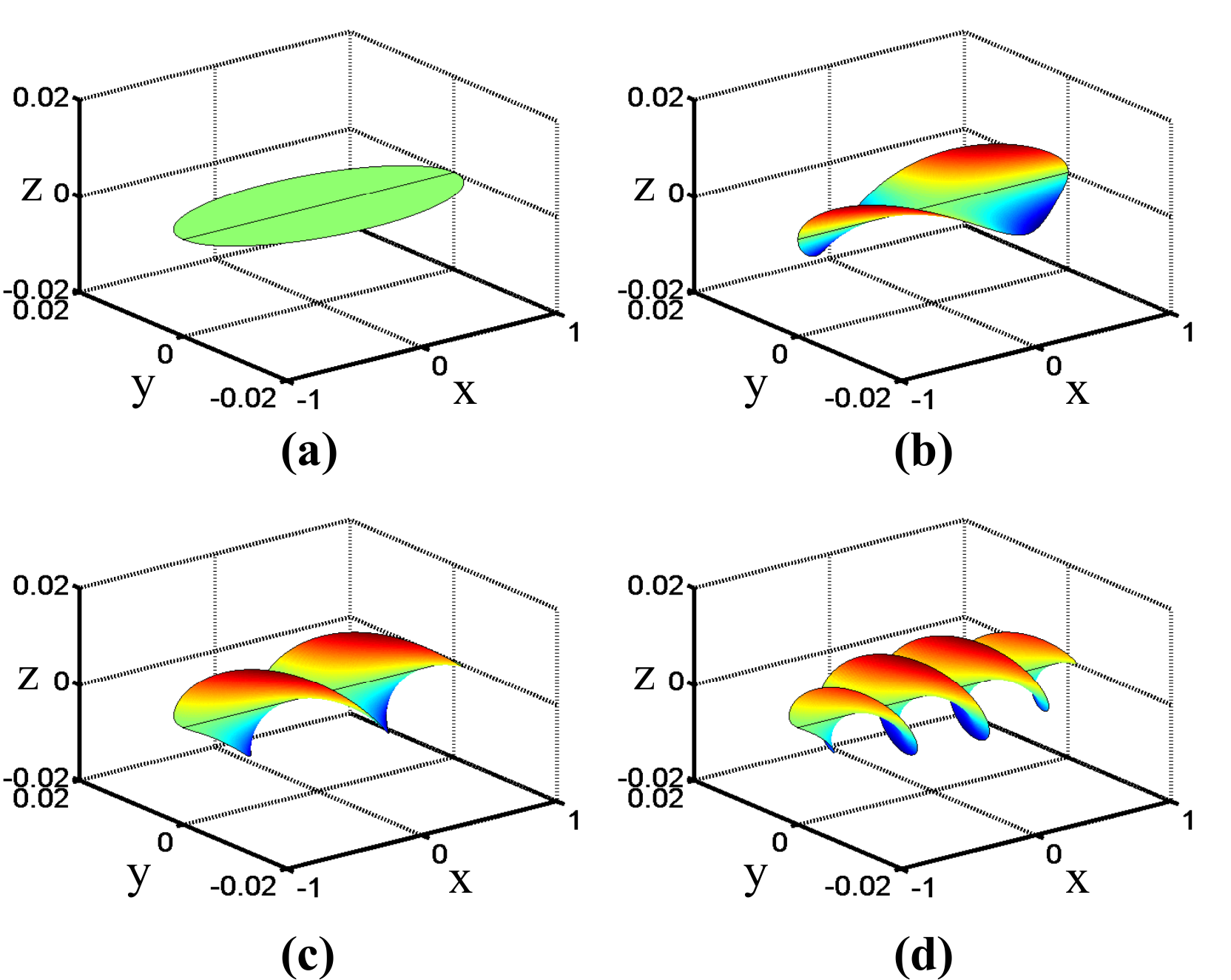}
\caption{Images of the ribbon plane for the twisted ellipsoidal swimmer with different number of twists, $Q$: (a) $Q=0$; (b) $Q=0.5$; (c) $Q=1$;  (d) $Q=2$. The different color shading corresponds to different heights in along the $z$ direction.}
\label{fig:twistellipse}
\end{center}
\end{figure}

The slender-ribbon formulation allows us to investigate  how twisting a ribbon changes its hydrodynamics. The behavior of a twisted plate ellipsoid, with aspect ratio $b_{\ell}=0.01$, is considered here in order to investigate this effect. The parameterization for such a body is given by
\begin{eqnarray}
\mathbf{r}(s_{1}) &=& s_{1}\mathbf{\hat{x}}, \\
\mathbf{\hat{T}} &=& \cos(Q \pi s_{1})\mathbf{\hat{y}}+\sin(Q \pi s_{1})\mathbf{\hat{z}},
\end{eqnarray}
where $Q$ determines the number of twists over the length of the body and $\mathbf{\hat{x}}$, $\mathbf{\hat{y}}$ and $\mathbf{\hat{z}}$ are orthogonal unit vectors. The ribbon planes for some of these shapes are illustrated in Fig.~\ref{fig:twistellipse}. 

\begin{figure}
\begin{center}
\includegraphics[width=\textwidth]{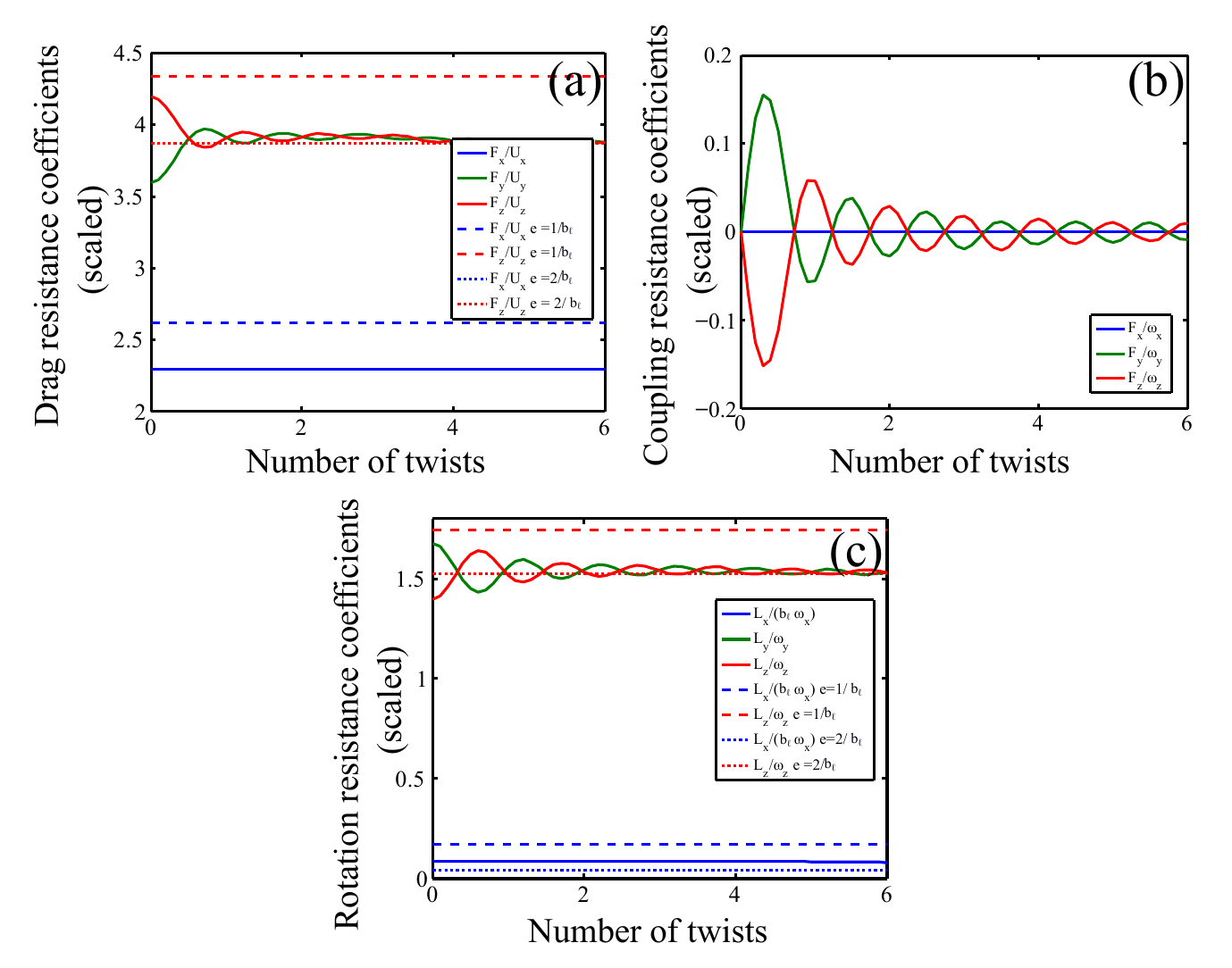}
\caption{Axial resistance coefficients of a twisted ellipsoid (computed in the center of mobility frame) as the number of twist, $Q$, increases: (a) Force from  translation; (b)  Force from angular rotation; (c) Torque form angular rotation. Solid lines are the SRT results while the dashed and dotted lines are the resistance coefficients for a prolate ellipsoid with aspect ratio $b_{\ell}^{-1}$ and $2 b_{\ell}^{-1}$ respectively. Numerical results for SRT were obtained using $N_{1}=20$ and $N_{2}=35$.}
\label{fig:twistellipseresult}
\end{center}
\end{figure}

We use SRT to compute the resistance matrix for such shapes. In Fig.~\ref{fig:twistellipseresult} we display  the values of the  matrix coefficients for the twisted ellipsoids as a function of the number of twists, $Q$. The calculations were carried out in the  center of resistance frame where the coupling matrices are symmetric. This frame is convenient as each of the sub-matrices, the force from translation, the force from rotation, and the torque from rotation, have only three terms. Also, wherever appropriate, the resistance coefficients for a prolate ellipsoid with aspect ratio 1/$b_{\ell}$ (dashed) and 2/$b_{\ell}$ (dotted) are plotted in Fig.~\ref{fig:twistellipseresult}.

We first observe that the components of the resistance matrix relating to force or torque parallel to the centerline remain constant as the number of twists increases while the other components display decaying oscillations. At leading order, the parallel force and torque are thus that of a plate ellipsoid and interactions between the twists are a higher-order effect.

The non-parallel terms are then seen to oscillate. This oscillating behavior can be attributed to which direction, $\mathbf{\hat{y}}$ or $\mathbf{\hat{z}}$, contains most of the ribbon plane. When more of the plane sits in $\mathbf{\hat{y}}$ the resistance to motion in $\mathbf{\hat{y}}$ is less than it is in $\mathbf{\hat{z}}$. At each $Q\approx 0.5$ interval equal amounts of the plane is in $\mathbf{\hat{y}}$ and $\mathbf{\hat{z}}$ and so there is no difference in the drag between these two dimensions. Further twisting again puts one direction ahead and so creates the observed oscillations. Note that the oscillations do not have an exact $Q=0.5$ periodicity because of the ellipsoidal cross section of the ribbon. The reduction in the oscillation amplitude is due the relative proportion of the helix plane producing the relative drag. Only the `extra ribbon' beyond half integer multiples contribute to the amplitude of the oscillation. As more twists occur over the  length of the ribbon, less of the total length of the ribbon is occupied by this `extra ribbon'. Therefore less of the ribbon is contributing to the difference in the resistances.

Eventually after enough twists one may suspect that the ribbon would begin to behave hydrodynamically like a prolate ellipsoid. The decreasing oscillations supports this idea.  Naively, one may expect that the limiting ellipsoid would have an aspect ratio of $b_{\ell}^{-1}$. 
However  Fig.~\ref{fig:twistellipseresult} shows that the resistance of the twisted ellipsoid is closer to a prolate spheroid with aspect ratios $2b_{\ell}^{-1}$ (dotted) than one of  aspect ratio $b_{\ell}^{-1}$ (dashed).  This factor of two could be considered as an average of the  dimensions of the ribbon along $\mathbf{\hat{T}}$ and $\mathbf{\hat{t}}\times\mathbf{\hat{T}}$.   It is possible that non-asymptotically thin ribbons would approach a prolate ellipsoid with aspect ratio $b_{\ell}^{-1}$ with a sufficient number of twists, however such behavior is of higher order in the asymptotic expansion.

Importantly, Fig.~\ref{fig:twistellipseresult}  shows that the hydrodynamic coefficient relating the force parallel to the tangent vector from  rotation around the tangent vector is zero at leading order for all values of $Q$. This result  appears counter-intuitive as the edges of the twisted ribbon trace out helices. Since the surface width of the sheet is of order $b_{\ell}$, and the rate of twisting along the length is assumed to be much less than $b_{\ell}^{-1}$ in the asymptotic expansion considered in this paper, the contribution to the flow by these helix edges is of order $b_{\ell}$ or smaller. Typically the absence of this coefficient would not be an issue as the  coupling created by any bending of the centerline would be an order of magnitude larger that that created by the twisting (see for example the twisted ribbon helix addressed above). However,  if the rate of twisting, $\sigma =|\partial_{s}\mathbf{\hat{T}}(s)|$, reaches order $b_{\ell}^{-1}$ ($Q \approx 30$), the force parallel to the tangent from rotations around the tangent vector would become significant. In said cases higher order terms in the SRT expansion would be needed. 

We finally note that past work considered similar twisted ribbons to investigate the diffusion of chiral objects in a shear flow \cite{Makino2005a}. The hydrodynamics  of these shapes was tackled  numerically using a boundary integral formulation and, though were asymptotically thin ($b\gg a$), they typically were very twisted, $\sigma = O(b_{\ell}^{-1})$, or very wide $\ell \sim b$. Although their ribbons  lie outside the asymptotic domain of  slender-ribbon theory,  the behavior reported  agrees with  the results described above \cite{Makino2005a}. The only resistance coefficients discussed in Ref.~\cite{Makino2005a} were the average rotational resistance perpendicular to the centerline tangent and the force from rotation parallel to the tangent. It was seen computationally that the average rotational resistance was close to that of a rod and had little dependence on the number of twists. Our results in 
Fig.~\ref{fig:twistellipseresult}c show that the resistance perpendicular to the centerline ($\mathbf{\hat{y}}$ and $\mathbf{\hat{z}}$) oscillate around the behavior of a rod and are out of phase with each other. The average of the two resistances is therefore close to a rod and does not change with the number of twists, similarly to the results of Ref.~\cite{Makino2005a}. Further, the value of the force from rotation parallel to the centerline was  shown in Ref.~\cite{Makino2005a} to go to 0 when the ribbons were straight or very very twisted. The authors commented in this work that ``The effect of the twist is generally weak in the principal part of the mobility tensor". This result supports our results that such dynamics is a higher order effect in the expansion of SRT.

\section{Conclusion and Outlook}

In this paper we have introduced ``slender-ribbon theory''  and computed asymptotically the leading-order hydrodynamics of a slender ribbon at low Reynolds number. 
The ribbon is represented by a plane of stokeslet singularities placed strictly inside the ribbon's body. The resulting kernel is asymptotically expanded in terms of two dimensionless groups: the ratio width over length, $b_{\ell} =b/\ell$, and the ratio thickness over  length, $a_{\ell}=a/\ell$. This expansion assumed that the curvature and the rate of twisting in the ribbon are less than $b_{\ell}^{-1}$. The resulting equations have many similarities to the  equations of slender-body theory  \cite{Johnson1979} and are seen to accurately determine all  resistance coefficients of a long flat ellipsoid. Unlike slender-body theory, no additional singularities beyond the stokeslet are required at leading order.

Slender-ribbon theory  was then used to characterize the behavior of different setups. First we investigated the dynamics of ribbons whose centerline are bent into a helix. The qualitative trends seen for thicker ribbons remained true in the asymptotically-thin limit.  
We  then  investigated the swimming hydrodynamics of an artificial microswimmer recently proposed experimentally, which  exploits the rotation of a helical ribbon to create propulsion. We obtained good  quantitative agreement between measurements and theoretical prediction of our asymptotic theory. Comparing slender ribbons to slender filaments it was  found that no equivalent slender body  can accurately replicate the dynamics of a helical slender ribbon. Finally an investigation into the dynamics of thin twisted ellipsoids showed that, as the  number of twists increased, the hydrodynamics limited towards that of an equivalent ellipsoid with a counter-intuitive  aspect ratio of $2 b_{\ell}^{-1}$. In that case, some of the resistance coefficients were seen to oscillate with an increase in twist, which was rationalized.

The asymptotic results derived in this paper could be used in  a number of other, more complex,  situations.  The dynamics of twisted planes opens up the possibility of exploring the hydrodynamics of topologically odd objects. Similarly, as slender-ribbon theory only requires the surface to move rigidly with a corresponding point on the ribbon plane,   the theory can handle non-rigid body motions, like waving, flapping and twisting.  In all these cases, the fluid flow around the body can also be computed, similarly to what is done in SBT \cite{Kim2004,Spagnolie2011}. This is achieved by plugging the computed force distribution into 
\begin{equation}
8 \pi \mathbf{u }(\mathbf{x}) =  \int_{-1}^{1} \,d t_{1} \rho_{1}(t_{1}) \int_{-1}^{1} \,d t_{2} \left[\frac{ \mathbf{f}(t_{1},t_{2})}{|\mathbf{R}'|} +\frac{ \mathbf{R}' \mathbf{R}'\cdot \mathbf{f}(t_{1},t_{2})}{|\mathbf{R}'|^{3}}\right],
\end{equation}
where $\mathbf{u }$ is the velocity of the fluid at $\mathbf{x}$ and $\mathbf{R}'$ is a vector from a point on the ribbon plane to $\mathbf{x}$, i.e.~$\mathbf{R}' =\mathbf{x}-  \mathbf{X}(t_{1},t_{2})$.

Slender-ribbon theory could  be further extended to include  extensions similar  to the ones developed  for slender bodies. This includes the interactions between multiple  bodies \cite{Tornberg2006,Kim2004}, the role of  nearby surfaces  \cite{Barta1988},  and the dynamics of  elastic shapes  \cite{Tornberg2004,Spagnolie2010,Spagnolie2011}.  
Furthermore, the practicality of slender-ribbon theory could  be increased by extending the domain of validity of  its derivation.  For example, extending the force density in the ribbon plane to higher order would allow to capture the resistance coefficient linking the force along the centerline tangent arising from  rotation around the tangent.  Another useful extension would be removing the mathematical assumption that $\mathbf{\hat{T}}\cdot\mathbf{\hat{t}}=0$ or allowing the ribbon to be a non-developable surface. This would allow for the parametrization to become similar to the ones used in solid mechanics  \cite{Dias2014}, setting the stage for the elasto-hydrodynamics of ribbons.

\section*{Acknowledgements}
The authors thank Eric Keaveny, Bradley J. Nelson, and Li Zhang for sharing their results and helpful discussion about their research. This research was funded in part by the European Union through a Marie Curie CIG Grant and the Cambridge Trusts. 

\appendix
\section{Asymptotic Integrals} \label{Appendix}
As discussed in Sec.~\ref{sec:SRE}, many of the integrals inside the stokeslet kernel take the form
\begin{equation}
\int_{-1}^{1} \,dt \frac{\chi^{i}}{\sqrt{\chi^{2} + \theta^{2}}^{j}}, \label{integral}
\end{equation}
where $\epsilon \chi = t -s$, $\theta$ is an arbitrary real function that does not depend on $t$ and $\epsilon$ is a small parameter. These integrals have been evaluated previously by Gotz \cite{Gotz2000} and  Table~\ref{tab:int} list all the relevant integrals of this form. In addition the following identities are also needed
\begin{equation}
\int_{-1}^{1} \,d t_{2} 
\log\left(\frac{1}{b_{\ell}^{2} \chi_{2,b_{\ell}}^{2} }\right) = 2(2-(1-s_{2})\log(1-s_{2}) -(1+s_{2})\log(1+s_{2})),
\end{equation}
and
\begin{equation}
\int_{-1}^{1} \,d t_{2} 
\log\left(\frac{1}{b_{\ell}^{2} \chi_{2,b_{\ell}}^{2}  + \frac{ a_{\ell}^{2} \rho_{2}^{2}}{b_{\ell}^{2} \rho_{1}^{2}}}\right) =   2(2-(1-s_{2})\log(1-s_{2}) -(1+s_{2})\log(1+s_{2}))  + 2 \pi \frac{a_{\ell} \rho_{2}}{b_{\ell} \rho_{1}} + O(a_{\ell}) + O(b_{\ell}^{2}).
\end{equation}

\begin{table}
\begin{tabular}{|c||c|c|c|c|}  \hline
j $\setminus$ i & 0 & 1 & 2 & 3 \\ \hline \hline
1 & $\,\,\epsilon \log\left( \frac{4(1-s^{2})}{\epsilon^{2} \theta^{2}}\right) + O (\epsilon^{3})$\,\, & $\,\,-2s + \frac{s \theta^{2} \epsilon^{2}}{1-s^{2}} +O (\epsilon^{3})$  \,\,&   &   \\ \hline
2 & $\frac{\epsilon \pi}{\theta} + \frac{2 \epsilon^{2}}{s^{2}-1} + O(\epsilon^{3})$ & $\,\,\epsilon \log\left(\frac{s-1}{s+1}\right) +O(\epsilon^{3})$\,\, & $2-\epsilon \pi \theta +O(\epsilon^{3})$ &  \\ \hline
3 & $\frac{2 \epsilon }{\theta^{2}} + O(\epsilon^{3})$ & $\frac{2 s \epsilon^{2}}{s^{2}-1}+ O(\epsilon^{3})$&$\,\,\epsilon \left[\log\left( \frac{4(1-s^{2})}{\epsilon^{2} \theta^{2}}\right) -2 \right] +O (\epsilon^{3})$\,\, &$\,\,-2s + \frac{3 s \theta^{2} \epsilon^{2}}{1-s^{2}} +O (\epsilon^{3})$ \,\, \\ \hline
4 & $\frac{\pi \epsilon}{2 \theta^{3}} + O(\epsilon^{3})$ &$ O(\epsilon^{3})$ & $\frac{\pi \epsilon}{2 \theta} + \frac{2 \epsilon^{2}}{s^{2}-1}+ O(\epsilon^{3})$ &$ \epsilon \log\left(\frac{s_{2}-1}{s_{2}+1}\right) + O(\epsilon^{3})$ \\ \hline
\end{tabular}
\caption{Table of the asymptotic solutions to the integral of the form of Eq.~\eqref{integral}}
\label{tab:int}
\end{table}

\bibliographystyle{ieeetr}
\bibliography{library}
\end{document}